\documentclass[twocolumn,tighten]{aastex631}

\usepackage{amsmath}
\usepackage{xcolor}
\usepackage{wasysym}
\usepackage{multirow}
\usepackage{newtxtext,newtxmath}
\usepackage{longtable}
\usepackage{afterpage} 
\usepackage[T1]{fontenc}
\usepackage{gensymb}
\usepackage{amsmath}
\usepackage{amssymb}

\usepackage{graphicx}	
\usepackage{amsmath}	
\usepackage{gensymb}    

\definecolor{my_color}{HTML}{3a18b1}

\definecolor{new_color}{HTML}{CF0000}
\definecolor{new_black}{HTML}{000000}
\usepackage{xfrac}
\definecolor{new_red}{HTML}{FF0000}

\newcommand\bedit[1]{\textcolor{new_black}{{#1}}}

\newcommand{\be}{\begin{equation}}
\newcommand{\ee}{\end{equation}}

\submitjournal{ApJL}

\shorttitle{TOI-5800}
\shortauthors{Jenkins et al.}

\begin{document}

\title{An Eccentric Sub-Neptune Moving Into the Evaporation Desert\footnote{This paper includes data gathered with the 6.5 meter Magellan Telescopes located at Las Campanas Observatory, Chile.}}

\author[0000-0001-9827-1463]{Sydney A. Jenkins} 
\altaffiliation{NSF Graduate Research Fellow}
\affiliation{Department of Physics, Massachusetts Institute of Technology, 77 Massachusetts Avenue, Cambridge, MA 02139, USA}
\affiliation{Kavli Institute for Astrophysics and Space Research, Massachusetts Institute of Technology, Cambridge, MA 02139, USA}
\author[0000-0001-7246-5438]{Andrew Vanderburg}
\affiliation{Department of Physics, Massachusetts Institute of Technology, 77 Massachusetts Avenue, Cambridge, MA 02139, USA}
\affiliation{Kavli Institute for Astrophysics and Space Research, Massachusetts Institute of Technology, Cambridge, MA 02139, USA}

\author[0000-0002-6576-3346]{Ritika Sethi}
\affiliation{Department of Physics, Massachusetts Institute of Technology, 77 Massachusetts Avenue, Cambridge, MA 02139, USA}
\affiliation{Kavli Institute for Astrophysics and Space Research, Massachusetts Institute of Technology, Cambridge, MA 02139, USA}

\author[0000-0003-3130-2282]{Sarah Millholland}
\affiliation{Department of Physics, Massachusetts Institute of Technology, 77 Massachusetts Avenue, Cambridge, MA 02139, USA}
\affiliation{Kavli Institute for Astrophysics and Space Research, Massachusetts Institute of Technology, Cambridge, MA 02139, USA}

\author[0000-0001-8812-0565]{Joseph E. Rodriguez}
\affiliation{Center for Data Intensive and Time Domain Astronomy, Department of Physics and Astronomy, Michigan State University, East Lansing, MI 48824, USA}


\author[0000-0003-4426-9530]{Luca Fossati}
\affiliation{Space Research Institute, Austrian Academy of Sciences, Schmiedlstrasse 6, A-8042 Graz, Austria}

\author[0000-0003-3615-4725]{Andreas Krenn}
\affiliation{Space Research Institute, Austrian Academy of Sciences, Schmiedlstrasse 6, A-8042 Graz, Austria}

\author[0000-0002-1533-9029]{Emily Pass}
\affiliation{Department of Physics, Massachusetts Institute of Technology, 77 Massachusetts Avenue, Cambridge, MA 02139, USA}
\affiliation{Kavli Institute for Astrophysics and Space Research, Massachusetts Institute of Technology, Cambridge, MA 02139, USA}

\author[0000-0002-8400-1646]{Alexander Venner}
\affiliation{Centre for Astrophysics, University of Southern Queensland, Toowoomba, QLD 4350, Australia}

\author[0000-0003-1305-3761]{R. Paul Butler}
 \affiliation{Earth and Planets Laboratory, Carnegie Institution for Science, 5241 Broad Branch Rd NW, Washington, DC 20015}

\author[0000-0002-4047-4724]{Hugh Osborn}
\affiliation{Center for Space and Habitability, University of Bern, Gesellschaftsstrasse 6, 3012 Bern, Switzerland} 
\affiliation{ETH Zurich, Department of Physics, Wolfgang-Pauli-Strasse 2, CH-8093 Zurich, Switzerland}

\author[0000-0002-5812-3236]{Aaron Householder}
\affiliation{Department of Earth, Atmospheric and Planetary Sciences, Massachusetts Institute of Technology, Cambridge, MA 02139, USA}
\affil{Kavli Institute for Astrophysics and Space Research, Massachusetts Institute of Technology, Cambridge, MA 02139, USA}

\author[0000-0002-0619-7639]{Carl Ziegler}
\affiliation{Department of Physics, Engineering and Astronomy, Stephen F. Austin State University, 1936 North St, Nacogdoches, TX 75962, USA}


\author[0000-0002-7733-4522]{Juliette Becker}
 \affiliation{Department of Astronomy, University of Wisconsin-Madison, 475 N. Charter Street, Madison, WI 53706, US}

\author{Perry Berlind}
\affiliation{Center for Astrophysics $|$ Harvard and Smithsonian, 60 Garden Street, Cambridge, MA 02138, USA}

 \author[0000-0001-6637-5401]{Allyson~Bieryla} 
\affiliation{Center for Astrophysics $|$ Harvard and Smithsonian, 60 Garden Street, Cambridge, MA 02138, USA}

\author[0000-0001-5132-2614]{Christopher Broeg}
 \affiliation{Space Research and Planetary Sciences, Physics Institute, University of Bern, Gesellschaftsstrasse 6, 3012 Bern, Switzerland}
 \affiliation{Center for Space and Habitability, University of Bern, Gesellschaftsstrasse 6, 3012 Bern, Switzerland}

\author[0000-0002-2830-5661]{Michael L. Calkins}
\affiliation{Center for Astrophysics $|$ Harvard and Smithsonian, 60 Garden Street, Cambridge, MA 02138, USA}

\author[0000-0002-5226-787X]{Jeffrey D. Crane}
\affiliation{The Observatories of the Carnegie Institution for Science, 813 Santa Barbara Street, Pasadena, CA 91101, USA}

\author[0000-0002-6939-9211]{Tansu Daylan}
\affiliation{Department of Physics and McDonnell Center for the Space Sciences, Washington University, St. Louis, MO 63130, USA}

 \author[0000-0003-2415-2191]{Julien de Wit}
\affiliation{Department of Earth, Atmospheric, and Planetary Sciences, Massachusetts Institute of Technology, 77 Massachusetts Avenue, Cambridge, MA 02139, USA}
\affiliation{Kavli Institute for Astrophysics and Space Research, Massachusetts Institute of Technology, Cambridge, MA 02139, USA}

\author[0000-0003-3773-5142]{Jason D.\ Eastman}
\affiliation{Center for Astrophysics \textbar \ Harvard \& Smithsonian, 60 Garden St, Cambridge, MA 02138, USA}

\author[0000-0001-9704-5405]{David Ehrenreich}
 \affiliation{Observatoire astronomique de l'Université de Genève, Chemin Pegasi 51, 1290 Versoix, Switzerland}
 \affiliation{Centre Vie dans l’Univers, Faculté des sciences, Université de Genève, Quai Ernest-Ansermet 30, 1211 Genève 4, Switzerland}

\author[0000-0002-9789-5474]{Gilbert A. Esquerdo}
\affiliation{Center for Astrophysics $|$ Harvard and Smithsonian, 60 Garden Street, Cambridge, MA 02138, USA}

\author[0000-0002-9113-7162]{Michael~Fausnaugh}
\affil{Department of Physics \& Astronomy, Texas Tech University, Lubbock TX, 79409-1051, USA}

\author[0000-0002-2036-2311]{Yadira Gaibor}
\affiliation{Department of Physics, Massachusetts Institute of Technology, 77 Massachusetts Avenue, Cambridge, MA 02139, USA}
\affiliation{Kavli Institute for Astrophysics and Space Research, Massachusetts Institute of Technology, Cambridge, MA 02139, USA}

\author[0000-0001-9911-7388]{David W. Latham}
\affiliation{Center for Astrophysics $|$ Harvard and Smithsonian, 60 Garden Street, Cambridge, MA 02138, USA}

\author[0000-0001-9699-1459]{Monika Lendl}
 \affiliation{Observatoire astronomique de l'Université de Genève, Chemin Pegasi 51, 1290 Versoix, Switzerland}

\author[0000-0002-7216-2135]{Andrew W. Mayo}
\affiliation{Department of Physics and Astronomy, San Francisco State University, San Francisco, CA 94132, USA}

\author[0000-0003-2029-0626]{Gaetano Scandariato}
\affiliation{INAF, Osservatorio Astrofisico di Catania, Via S. Sofia 78, 95123 Catania, Italy}

\author[0000-0002-8681-6136]{Steve Shectman}
\affiliation{The Observatories of the Carnegie Institution for Science, 813 Santa Barbara Street, Pasadena, CA 91101, USA}

\author[0009-0008-5145-0446]{Stephanie Striegel}
\affiliation{SETI Institute, Mountain View, CA 94043 USA/NASA Ames Research Center, Moffett Field, CA 94035 USA}

\author[0009-0008-2801-5040]{Johanna Teske}
 \affiliation{Earth and Planets Laboratory, Carnegie Institution for Science, 5241 Broad Branch Rd NW, Washington, DC 20015}
\affiliation{The Observatories of the Carnegie Institution for Science, 813 Santa Barbara Street, Pasadena, CA 91101, USA}

\author[0000-0001-8749-1962]{Thomas G Wilson}
 \affiliation{Department of Physics, University of Warwick, Gibbet Hill Road, Coventry CV4 7AL, United Kingdom}


\correspondingauthor{Sydney Jenkins}
\email{sydneyaj@mit.edu}



\begin{abstract}
Though missions such as \textit{Kepler}, \textit{K2}, and \textit{TESS} have discovered >2,000 sub-Neptune and Neptunian planets, there is a dearth of such planets at close-in (P$\lesssim$3 days) orbits. This feature, called the Neptune desert or the evaporation desert, is believed to be primarily shaped by planetary migration and photoevaporation. However, this region is not completely devoid of planets -- a small number of very hot Neptunes reside within the desert. These planets provide an opportunity to directly probe the effects of migration and photoevaporation. We present confirmation of TOI-5800 b, an eccentric sub-Neptune on a $\approx$2.6 day period that is likely actively undergoing tidal migration. We use radial velocity measurements from the Carnegie Planet Finder Spectrograph (PFS) to constrain TOI-5800 b’s mass and eccentricity. We find that it has an unusually high eccentricity (0.39$\pm$0.07) for its short orbit. TOI-5800 is therefore currently experiencing high levels of tidal heating as it moves into the desert. Ranked as a top candidate for transmission and emission spectroscopy within its temperature and radius regime, TOI-5800 b is a prime target for atmospheric characterization with JWST. TOI-5800 b presents a unique opportunity to study the atmosphere of a planet undergoing tidal heating and to probe the composition of sub-Neptune planets.
\end{abstract}



\section{Introduction}

\begin{figure*}
\centering
\includegraphics[width=.95\textwidth]{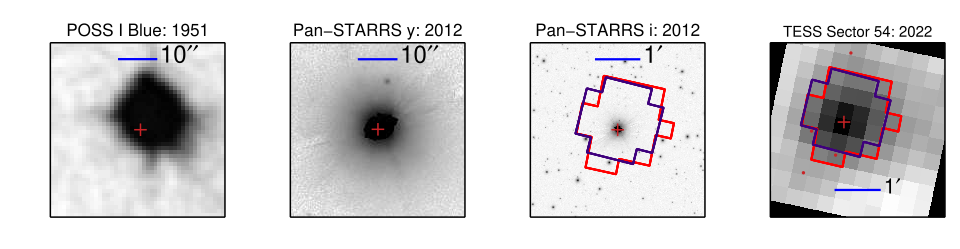}
\caption{Archival imaging of TOI-5800. From left to right, images are from Palomar Observatory Sky
Survey \citep{abe1955}, the Panoramic Survey
Telescope and Rapid Response System \citep[Pan-STARRS,][]{pan2016} survey in the $y$ band, the Pan-STARRS survey in the $i$ band, and co-added \textit{TESS} images from Sector 54. In all images, we show the current position of TOI-5800 with a red cross. The photometric apertures for Sectors 54 and 81 are shown in red and purple, respectively. In the rightmost image, nearby stars  within 6 magnitudes \citep{gai2016, gai2023} of TOI-5800 are shown as red points. \label{TESS_image}}
\end{figure*}

Hot sub-Neptunes, which have short-periods ($\lesssim$ 100 days) and radii roughly 2-4 times that of Earth, are known to be common in our galaxy, but are missing from our own solar system. As a result, their evolutionary history and physical properties are relatively poorly understood. In the past fifteen years, however, missions such as \textit{Kepler}, \textit{K2}, and \textit{TESS} have discovered thousands of sub-Neptunes, making them the subject of intensive speculation and study.

One important clue about the physical properties of sub-Neptunes and the processes that sculpt their population is the fact that there is a dearth of these planets at close-in (P$\lesssim$3 days) orbits. This feature, called the ``Neptune desert'' or the ``evaporation desert,'' is believed to be primarily shaped by high eccentricity migration (HEM) and photoevaporation \citep[e.g.,][]{2018Natur.553..477B, 2018MNRAS.479.5012O}.  In this scenario, 
a massive companion drives the Neptunian planet to high eccentricities via planet-planet scattering or Lidov-Kozai cycles. This mechanism has been explored for systems such as HAT-P-11 b \citep{lu2024} and WASP-107 \citep{yu2024}. As the perturbed planet undergoes tidal circularization, its orbit shrinks, driving the planet into the desert. At this new, close-in orbit, the planet may then lose a significant amount of mass via atmospheric evaporation. If sufficient mass is lost, the planet becomes a smaller, denser super-Earth. 
This hypothesis is corroborated by the many warm Neptunes that exhibit high rates of mass loss, including GJ 436 b \citep{2015Natur.522..459E}, GJ 3470 b \citep{2018A&A...620A.147B}, HAT-P-11 b \citep{2022NatAs...6..141B}, and HAT-P-26 b \citep{2022AJ....164..234V}.

Though \textit{Kepler} found few planets in the evaporation desert \citep{lun2016, maz2016}, \textit{TESS} has recently found a number of examples \citep[e.g.,][]{Jenkins2020, Arm2020, Mur2021, Esp2022, Mor2022, Per2022, Lil2023, Nabbie2024}.  These planets provide an opportunity to directly probe the effects of migration and photoevaporation. Of these known sub-Neptunes, only 18 have an eccentricity $>3\sigma$ above zero \citep[NASA Exoplanet Archive,][]{ake2013}, making them possible examples of planets actively entering the desert via high-eccentricity migration \bedit{(HEM)}. Identifying and characterizing additional eccentric Neptunes in the desert would provide critical insight into how the desert is formed. Additionally, because the leading explanation for the desert assumes that most Neptunes have a hydrogen- or helium-rich envelope that can be easily lost in high irradiation environments, understanding how planets enter the desert also has important implications for the composition of Neptunian worlds.

In this work, we characterize TOI-5800 b, a high-eccentricity \textit{TESS} planet candidate discovered in 2022 with a period of $\approx$2.6 days. Though TOI-5800 b was originally classified as a possible false positive \citep{hor2023}, we combine photometric and spectroscopic observations of TOI-5800 with RV measurements to confirm that it is a planet and to provide improved constraints on the system. The paper is organized as follows: we describe the \textit{TESS} photometry and follow-up photometric and spectroscopic observations in \S\ref{Observations}. We then present revised stellar parameters in \S\ref{Stellar_parameters} and outline our transit and RV fitting routine in \S\ref{Modeling}. We discuss constraints on TOI-5800 b's density and eccentricity in \S\ref{Results}, in addition to constraints on the presence of any planetary companions. We then explore possible explanations for the planet's high eccentricity, and implications for the planet's dynamical state, in \S\ref{State}. We summarize the planet's dynamical history, composition, and prospects for future atmospheric characterization in \S\ref{Discussion}, before concluding in \S\ref{Conclusion}.

\section{Observations}
\label{Observations}
\subsection{TESS}
\label{TESS}
TOI-5800 b was first identified as a planet candidate by \textit{TESS}. It was observed in two sectors (54 and 81) that began in July 2022 and July 2024. It was observed at 2-minute and 20-s cadences in the first and second sectors, respectively. The photometric data was then processed by the Science Processing Operations Center \citep[SPOC,][]{jen2016} pipeline at NASA Ames Research Center. SPOC produced a threshold-crossing event (TCE) that was fit with an initial limb-darkened transit model \citep{li2019} and passed a series of diagnostic tests \citep{twi2018}. The \textit{TESS} Science Office (TSO) reviewed the results from sector 54 and issued an alert on 2022 September 22. The difference image centroids analysis indicates that the host star is within 5.5$\pm$2.7'' and 4.1$\pm$2.7'' of the source of the transit signal in sectors 54 and 81, respectively. The aperture masks used for simple aperture photometry \citep[SAP,][]{twi2010,mor2020} are shown in Figure~\ref{TESS_image}. One star overlaps the aperture, but is sufficiently faint ($>$5 $Gaia$ magnitude difference) that we do not expect it to dilute the \textit{TESS} light curve. 

We noticed that several transits were excluded from the default \textit{TESS} light curves because of high levels of scattered light in the background of the \textit{TESS} images. Because TOI-5800 is a bright star, we were able to recover useful data during some of these times by performing our own reduction. We started from the calibrated target pixel files and extracted light curves following \citet{Vanderburg2016ApJS}. We derived systematics corrections following \citet{Vanderburg2019ApJL} and performing a linear least squares fit to the SAP light curve. We modeled the light curve as the sum of a basis spline (to account for stellar variations and long-timescale instrumental systematics), a time series of the background counts, several vectors describing moments of the spacecraft quaternion time series within each exposure, and vectors derived by the SPOC pipeline (specifically the band 3 cotrending basis vectors from the Pre-search Data Conditioning pipeline, \citealt{smi2012, stu2012, stu2014}). We performed a linear least squares fit via matrix inversion, ignoring points taken during transit in the fit and iteratively excluding any other outlier points in the light curve. After producing the systematics-corrected light curve, we removed low-frequency variations by simultaneously fitting a basis spline with a transit model \citep[similar to][but without the simultaneous fit to spacecraft systematics]{Vanderburg2016ApJS}. We use this flattened light curve, shown in Figure~\ref{Transits}, in our modeling described in Section \ref{Modeling}.



\subsection{Reconnaissance Observations}

\subsubsection{Spectroscopy} 
Reconnaisance spectra of TOI-5800 were obtained as part of the \textit{TESS} Follow-Up Observing Program (TFOP)\footnote{\url{https://tess.mit.edu/followup/}} SubGroup 2 (SG2) Reconaissance Spectroscopy program. We collected eight spectra between 2022 October 13 and 2023 December 18 with the Tillinghast Reflector Echelle Spectrograph \citep[TRES,][]{gaborthesis}, an echelle spectrograph on the 1.5-meter Tillinghast telescope that covers wavelengths between 385 to 910 nm. The average SNR per resolution element is 40.0 near the magnesium triplet. The TRES spectra are used to help constrain the stellar properties (see \S\ref{spectroscopic_parameters}). 



\subsubsection{High Resolution Imaging} 
If there are stars visually close to TOI-5800, their flux contributions may make it more difficult to detect shallow transit events. This contamination could also make the radius of any detected planets appear smaller, leading to an overestimated planetary density. High-resolution imaging is therefore critical for ruling out the presence of any nearby stars. This follow-up was performed as part of the TFOP SubGroup 3 (SG3) High-Resolution Imaging program. We obtained observations with 
the High-Resolution Camera \citep[HRCam, ][]{tok2008, Tok2018, zie2020} at the SOAR 4.1-m telescope on 2023 August 31. HRCam uses speckle interferometry to search for nearby stars. We collected 400 25 ms frames, and processed the data as described in \citet{Tok2018}. 
No additional sources were found within 2.12 magnitudes at separations $>$ 0.1'', or within 5.85 magnitudes at separations $>$ 1''.

\begin{figure*}
\centering
\includegraphics[width=.95\textwidth]{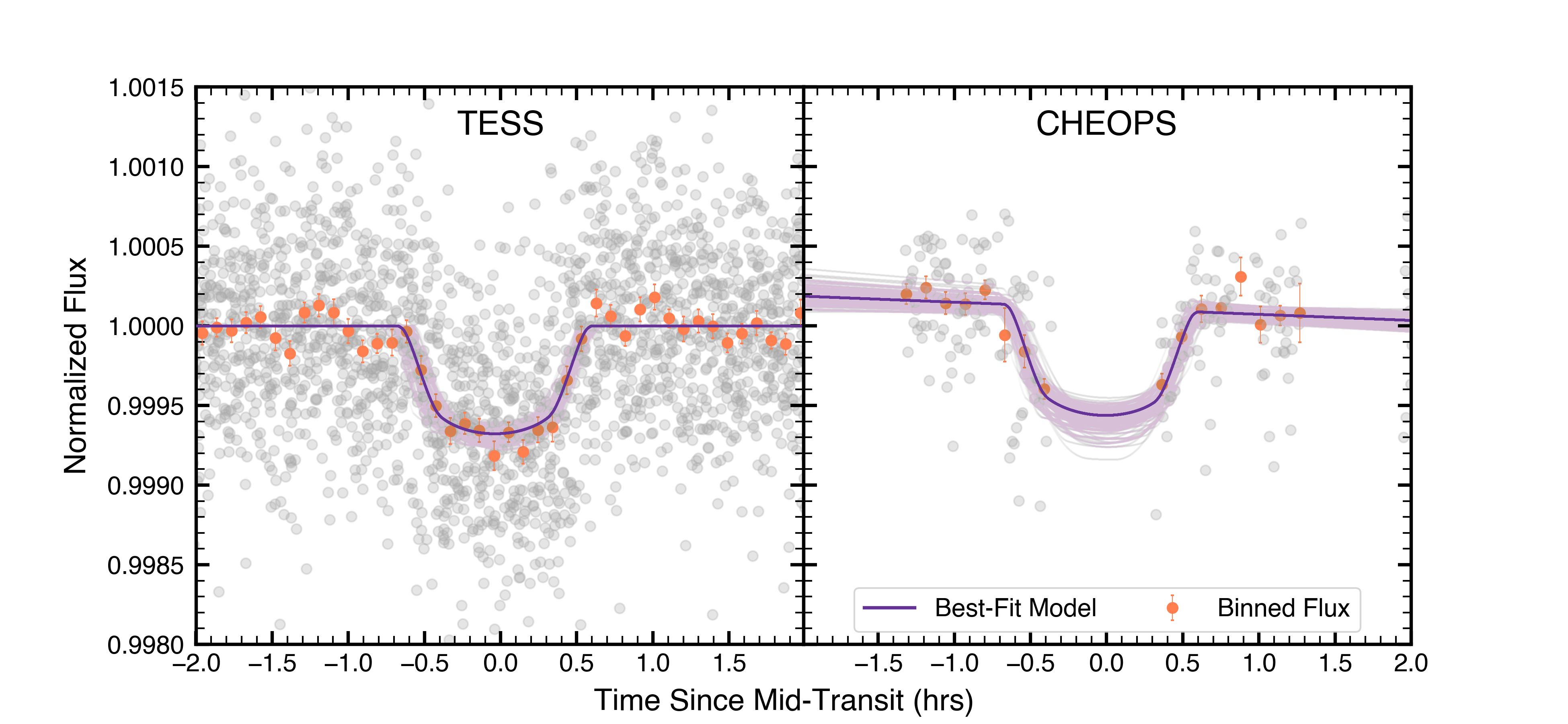}
\caption{Photometric data from \textit{TESS} and CHEOPS (see \S\ref{TESS} and \ref{CHEOPS}), shown in gray. Binned data is shown in orange, best-fit models (see \S\ref{Modeling}) are shown in dark purple, and 100 draws from the posterior distribution are shown in light purple.  \label{Transits}}
\end{figure*}

\subsection{CHEOPS}
\label{CHEOPS}
TOI-5800 was observed by the CHaracterising ExOPlanet Satellite (CHEOPS) satellite in 2023 and 2024 (PR149000\_TG021801 \& PR149000\_TG002001). CHEOPS' payload comprises a high-precision photometric camera fed by a 30 cm telescope \citep{Benz2021}. Observations were taken as part of the CHEOPS GTO programme “Chateaux” (CHeops And TEss vAlidate Unconfirmed eXoplanets, CH\_PR9000 PI Osborn), which is part of CHEOPS’ synergy with other missions. In order to produce precise photometry, we initially used the PSF fitting tool PSF Imagette Photometric Extraction \citep[PIPE,][]{Morris2021, Brandeker2024}\footnote{\url{https://github.com/alphapsa/PIPE}}. Due to the field rotation, photometry must typically be corrected for systematic effects to approach the expected photometric precision. We decorrelated the photometry using the {\tt chexoplanet} package \citep[e.g.,][]{Egger2024}\footnote{\url{https://github.com/hposborn/chexoplanet}}. {\tt chexoplanet} assesses various housekeeping data timeseries, such as onboard centroids, temperature, background, and vectors extracted from the residuals of the PSF photometry, to determine which improve the photometric timeseries. This decorrelation was performed on both observations simultaneously, alongside a cubic bspline model of the roll angle with 9-degree spacing to remove rapid variations in flux with spacecraft orientation, and a transit model of the planet to avoid overfitting the astrophysical data. Unfortunately, in the case of TOI-5800b, interruptions due to Earth occultations occurred during mid-transit for both CHEOPS visits, though in- and egress is seen in both visits. The result was photometry with photometric precision of 33 ppm hr$^{-1}$.

\subsection{High-Resolution Spectroscopy with PFS}
\label{PFS}
We obtained high-precision RV measurements of TOI-5800 with the Planet Finder Spectrograph \citep[PFS,][]{cra2008,cra2010} on the 6.5-m Magellan II Telescope. PFS is a high-precision echelle spectrograph with a resolution up to $\sim$130,000 in the 1$\times$2 binning mode. We observed the TOI-5800 system 25 times with PFS between May and October 2024. All observations were 900 seconds long, with an average uncertainty of 0.83 m s$^{-1}$. 
The spectra were reduced with the standard PFS reduction pipeline \citep{but1996, cra2006}. We give all RV measurements and corresponding uncertainties in Table~\ref{tab:RVs} and Figure~\ref{RV}.


\section{Stellar Characterization} 
\label{Stellar_parameters}

\subsection{Spectroscopic Parameters}
\label{spectroscopic_parameters}
We first characterize TOI-5800 using the SG2 TRES data and the Stellar Parameter Classification (SPC) tool \citep{buc2012}. SPC cross-correlates an observed spectrum against a grid of model template spectra. The model spectra, calculated using the \citet{kur1992} grid of model atmospheres, cover the wavelength region between 5050 and 5360 \AA\, and correspond to varying effective temperature $T_{\textrm{eff}}$, surface gravity log($g$), metallicity [m/H], and projected equatorial rotational velocity $V_{\textrm{rot}}$. Yonsei-Yale isochrones \citep{Yi2001} are used to set priors on log($g$). The synthetic spectra span the following ranges: 3500 K $< T_\textrm{eff} <$  9750 K, 0.0 $<$ log($g$) $<$ 5.0, $-$2.5 $<$ [m/H] $<$ +0.5, and 0 km s$^{-1}$ $< V_{\textrm{rot}} <$ 200 km s$^{-1}$. Using SPC, we estimate $T_{\textrm{eff}}=$4830$\pm$50 K, log($g$)$=$4.4$\pm$0.1, and [m/H]$=$0.02$\pm$0.08 dex. 

\subsection{Fundamental Parameters}
We then use {\tt astroARIADNE} \citep{vin2022} to estimate the host star's mass and radius. This Python package fits broadband photometry to the {\tt Phoenix v2} \citep{hus2013}, {\tt BtSettl} \citep{all2012}, \citet{cas2003}, and \citet{kur1993} stellar atmosphere models. It then uses MESA Isochrones and Stellar Tracks \citep[MIST,][]{2011ApJS..192....3P, 2013ApJS..208....4P, pax15, dot2016, cho2016} isochrones to estimate the stellar mass. We apply Gaussian priors to $T_{\textrm{eff}}$ and [m/H] centered on the values previously estimated with SPC. We include the bandpasses $G$, $G_{BP}$, $G_{RP}$ (\textit{Gaia} DR2), WISE (Wide-field Infrared Survey Explorer) W1-W2, and $J$,$H$,$K_s$ (2MASS, Two Micron All Sky Survey), in addition to GALEX NUV, the Johnson $B$ and $V$ magnitudes (APASS, AAVSO Photometric All-Sky Survey), and the Tycho $B$ and $V$ magnitudes. Though {\tt astroARIADNE} automatically uses \textit{Gaia} DR2, we note that the photometric values are consistent with \textit{Gaia} DR3. All photometric data are given in Table~\ref{tab:Stellar_properties}. We apply the systematic uncertainty floors used by {\tt EXOFASTv2} \citep{eas2019}. 
We estimate a radius and mass of 0.80$\pm$0.01 R$_{\star}$ and 0.79$^{+0.05}_{-0.02}$ M$_{\star}$, respectively, corresponding to a stellar density of 2.18$^{+0.14}_{-0.08}$ g cm$^{-3}$. The luminosity is 0.31$\pm$0.01 L$_{\star}$, and the extinction is close to zero \citep[$E$($g-r$) = 0.05$^{+0.01}_{-0.02}$ mag, ][]{Green_2019}. 

We perform a parallel analysis of the stellar parameters using {\tt EXOFASTv2} as a cross-check of our results. The estimated stellar parameters are consistent with the values found using {\tt astroARIADNE}. 

\subsection{Stellar Age}
\label{Stellar Age}
We constrain the system's age using the TRES spectra. We first calculate the stellar activity indicator $R^\prime_{HK}$ for each spectrum (as described in \citealt{2016ApJ...829L...9V,2018AJ....155..136M}). Though we expect $R^\prime_{HK}$ to vary over time, the average value can still provide insight into the star's level of activity. We measure \bedit{a weighted average} of \bedit{$-$4.75 $\pm$ 0.07.} We then use the activity-age relation from \citet{mam2008} to estimate a stellar age of $\approx$2.5 Gyr. The formal uncertainties on this estimate are about \bedit{30-40\% (2.5$^{+1.0}_{-0.7}$ Gyr)}, but there are likely systematic effects that increase this uncertainty. This is consistent with several other age measurements for the system. For instance, we do not detect a strong rotation period, though we expect the period to be $>$18$\pm$4 days based on TRES $v$sin$i$ measurements. This suggests that the star is at least several hundred Myr old. Additionally, we estimate the system's 3D kinematics to be (U,V,W)$=$($-$16.09$\pm$0.29, $-$10.92$\pm$0.43, 5.70$\pm$0.26) km s$^{-1}$ with respect to the Local Standard of Rest \citep[LSR,][]{cos2011}, corresponding to a 99.0\% probability of belonging to the thin disk \citep{ben2014}. This suggests that the star is also not extremely old, as thin disk stars have a roughly uniform age distribution with an upper limit of $\sim$8$-$10 Gyr \citep[e.g.,][]{ben2003, hay2013, Kil2017, xia2017}. In comparison, thick disk stars have an average age of $\sim$9$-$10 Gyr \citep{Kil2017, sha2019}. Our result is, however, mildly inconsistent with the value predicted by {\tt astroARIADNE}, which uses MIST isochrones to estimate an age of 11.6$^{+1.2}_{-7.3}$ Gyr. We note that ages derived using isochrones are known to have large systematic uncertainties \citep{tor2009}, and that the estimated age distribution covers most of the age of the universe.


\begin{table}
\centering
\caption{TOI-5800 Stellar Properties.} \label{tab:Stellar_properties}
\centering
\begin{tabular}{lll}
\hline
Parameter & Value & Source \\
\hline
\multicolumn{3}{c}{Astrometry} \\ 
\hline
RA (\degree) & 305.065292 & $Gaia$ DR3 \\
Dec (\degree) & $-$7.411697 & $Gaia$ DR3 \\
$\mu _{\alpha}$cos$\delta$ (mas yr$^{-1}$) & 51.531 $\pm$ 0.018  & $Gaia$ DR3 \\
$\mu _{\delta}$ (mas yr$^{-1}$) & $-$52.722 $\pm$ 0.015 & $Gaia$ DR3 \\
$\pi$ (mas) & 23.395 $\pm$ 0.019 & $Gaia$ DR3 \\

\hline
\multicolumn{3}{c}{Photometry} \\ 
\hline
$G$ (mag) & 9.2255 $\pm$ 0.0028 & $Gaia$ DR2 \\
$G_{\mathrm{BP}}$ (mag) & 9.7396 $\pm$ 0.0028  & $Gaia$ DR2 \\
$G_{\mathrm{RP}}$ (mag) & 8.5544 $\pm$ 0.0038 & $Gaia$ DR2 \\
$W1$ (mag) & 7.111 $\pm$ 0.044 & AllWISE \\
$W2$ (mag)  & 7.188 $\pm$ 0.021 & AllWISE \\
$J$ (mag)  & 7.771 $\pm$ 0.021 & 2MASS \\
$H$ (mag) & 7.284 $\pm$ 0.026 & 2MASS \\
$K_s$ (mag)  & 7.191 $\pm$ 0.020 & 2MASS \\
NUV (mag) & 17.292 $\pm$ 0.031 & GALEX \\
$B_{\mathrm{J}}$ (mag) & 10.484 $\pm$ 0.057 & APASS \\
$V_{\mathrm{J}}$ (mag) & 9.534 $\pm$ 0.030 & APASS \\
$B_{\mathrm{T}}$ (mag) & 10.745 $\pm$ 0.048 & Tycho-2 \\
$V_{\mathrm{T}}$ (mag) & 9.658 $\pm$ 0.028 & Tycho-2 \\

\hline
\multicolumn{3}{c}{Stellar Parameters} \\ 
\hline
Spectral type & K3V & This work$^*$ \\
T$_{eff}$ (K) & 4830 $\pm$ 50 & This work \\
 $\left[\mathrm{m}/\mathrm{H}\right]$ (dex) & 0.02 $\pm$ 0.08 & This work \\
 L (L$_{\odot}$) & 0.31$\pm$0.01 & This work \\
R (R$_{\odot}$) & 0.80 $\pm$ 0.01 & This work \\
M (M$_{\odot}$) &  0.79$^{+0.05}_{-0.02}$ & This work \\
$\rho_{\star}$ (g cm$^{-3}$) & 2.18$^{+0.14}_{-0.08}$ & This work \\
log$R$'$_{\mathrm{HK}}$  & \bedit{$-$4.75 $\pm$ 0.07} & This work \\
Age (Gyr) & \bedit{2.5$^{+1.0}_{-0.7}$} & This work \\
\hline
\multicolumn{3}{l}{$GAIA$ \citep{gai2016, Gaia2018, gai2023};}\\
\multicolumn{3}{l}{AllWISE \citep{wri2010}; 2MASS \citep{skr2006};}\\
\multicolumn{3}{l}{GALEX \citep{bia2011}; APASS \citep{hen2014};}\\
\multicolumn{3}{l}{Tycho-2 \citep{hog2000}.}\\
\multicolumn{3}{l}{$^*$Using \citet{pec2013}.}\\
\end{tabular}

\end{table}

\begin{figure*}
\centering
\includegraphics[width=.95\textwidth]{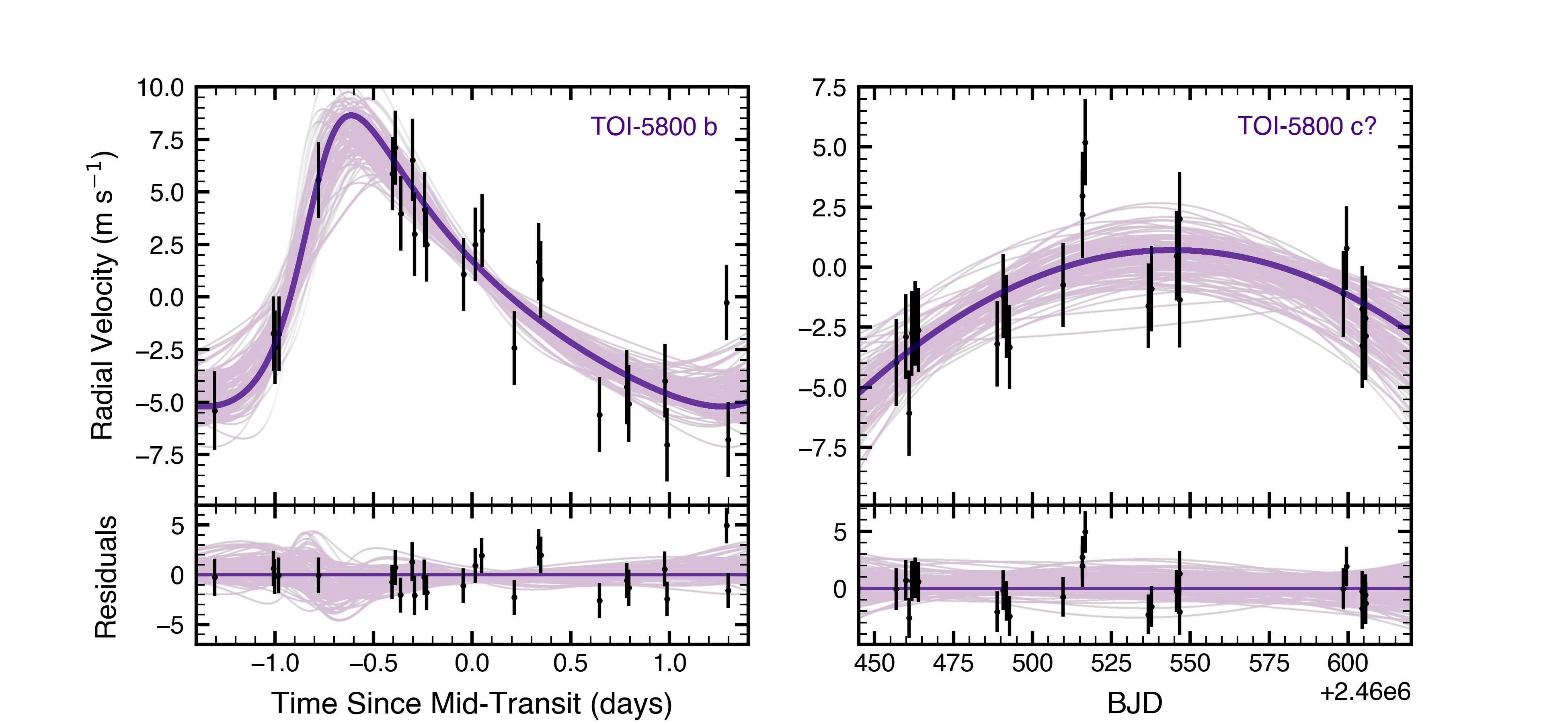}
\caption{Radial velocity data from PFS (see \S\ref{TESS}), shown in black. The best-fit model (see \S\ref{Modeling}) is shown in dark purple, and 100 draws from the posterior distribution are shown in light purple. Left: Phase-folded data and model with quadratic component subtracted. Right: Quadratic fit. We find tentative evidence for a quadratic trend in the RV data, potential evidence of an outer planetary companion.  \label{RV}}
\end{figure*}

\section{Analysis and Modeling}
\label{Modeling}
We determine best-fit parameters and uncertainties for the TOI-5800 system by performing a custom analysis of the transit photometry and radial velocity observations. 
 We use the Python package {\tt batman} \citep{kre2015} to calculate the model transit light curves and the {\tt radvel} \citep{ful2018} package to calculate the model RV curves. We fit the planet mass M$_P$, planet-to-star radius ratio R$_P/$R$_{\star}$, time of inferior conjunction $T_0$, semi-major axis $a/$R$_{\star}$, and inclination $\cos{i}$. We also fit the eccentricity $e$ and argument of periastron $\omega$ using the parameterization $\sqrt{e}\cos{\omega}$ and $\sqrt{e}\sin{\omega}$. We additionally fit a jitter term for each instrument (\textit{TESS}, CHEOPS, and PFS), which is added in quadrature to the reported uncertainties in order to account for any additional astrophysical or instrumental uncertainties. We also include an offset term for each instrument, two quadratic limb darkening terms for both the \textit{TESS} and CHEOPS observations, and a linear slope for the CHEOPS observations. In total, our likelihood function includes 19 free parameters. \bedit{All parameters and corresponding priors are given in Table~\ref{tab:Planet_properties}.}

We require the impact parameter to be less than 1$+$R$_P/$R$_{\star}$ and the perihelion distance to be larger than the stellar radius. These restrictions constrain the sampler to planets that are transiting and not engulfed by their star. We additionally set Gaussian priors on the stellar density and on the four limb darkening coefficients. The stellar density prior is centered on the value given in Table~\ref{tab:Stellar_properties}, with a standard deviation corresponding to the measured uncertainty. To determine the limb darkening priors, we estimate the limb darkening coefficients using the {\tt ldtk} \citep{par2015} package. {\tt ldtk} uses the {\tt Phoenix v2} model stellar spectra and applies constraints based on the estimated stellar $T_{\textrm{eff}}$, log($g$), and [m/H]. We find $TESS$ coefficients of 0.52 and 0.10, and CHEOPS coefficients of 0.65 and 0.06. We center the limb darkening priors on these values, with a standard deviation of 0.20. 


We explore the parameter space with a differential evolution Markov Chain Monte Carlo (MCMC) sampler. Sampling \bedit{is} done using the {\tt edmcmc} package \citep{andrew_vanderburg_2021_5599854}. We evolve 128 chains for 80,000 iterations each \bedit{and} test for convergence by calculating the Gelman-Rubin statistic for each parameter. All values are below 1.1, indicating that our model is successfully converging. \bedit{For parameters with uniform priors, we also verify that the priors do not affect the posterior distribution.} The best-fit parameter values are given in \bedit{Table~\ref{tab:Planet_properties}}, and the final transit and RV models are shown in Figures~\ref{Transits} and~\ref{RV}.

Again, as a cross check on the results of our modeling, we also run an independent analysis of the entire system with {\tt EXOFASTv2} \citep{eas2019}. We use the same combined \textit{TESS}, CHEOPS, and PFS data used in the previously described MCMC analysis. The results are consistent within $\sim1\sigma$ for all parameters. \bedit{We also check whether there are significant transit timing variations (TTVs) that could be skewing our posterior inferences. We measure the individual transit times from the \textit{TESS} dataset and show the resulting TTVs in Figure~\ref{TTVs}. We do not identify any significant signals in the transit times. } 


\section{Results}
\label{Results}
\subsection{Confirmation of Planet TOI-5800 b}
\begin{figure}
\centering
\includegraphics[width=\columnwidth]{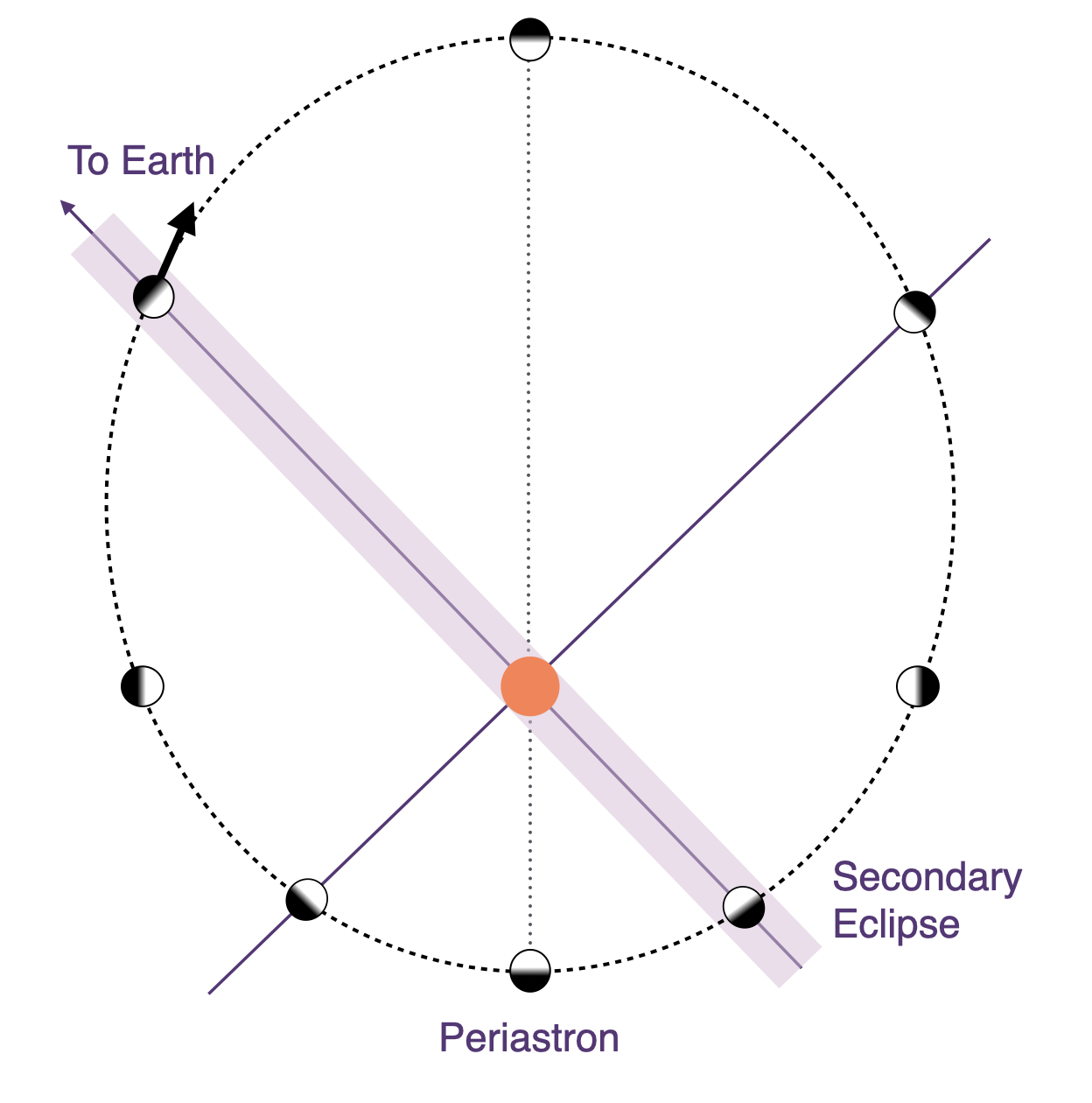}
\caption{Diagram of TOI-5800 system. The star is shown to scale with the planet's orbit.  \label{Diagram}}
\end{figure}

The primary result of our work is the confirmation of TOI-5800 b as a genuine exoplanet. Although TOI-5800 b had previously been classified as a potential false positive \citep{hor2023} based on a statistical analysis of the transit photometry, our radial velocity observations confirm that it is a planet. In a periodogram search of the PFS RV observations, we detect a signal at the orbital period originally identified by \textit{TESS}. The fact that we independently detect the planet's signature in both the \textit{TESS} and PFS data is ironclad confirmation of TOI-5800 b's planetary nature and allows us to  reject all false positive scenarios. We show a diagram of the system in Figure~\ref{Diagram}.

\subsection{Constraints on Additional Planetary Companions}
\label{Companion}

\bedit{We additionally} find tentative evidence for an outer companion in the RV data. The maximum residual between the data and the one-planet model is 6.36 m s$^{-1}$, with a root mean square of 2.35 m s$^{-1}$. When we include a quadratic trend in our fit, we find a best-fit RV slope and quadratic term of 0.014 $\pm$ 0.009 m s$^{-1}$ days$^{-1}$ and $-$0.0007 $\pm$ 0.0002 m s$^{-1}$ days$^{-2}$, respectively. The RV quadratic term is therefore $>3\sigma$ from 0 m s$^{-1}$ days$^{-2}$. This quadratic trend is shown in  Figure~\ref{RV}. We use the best-fit values from this fit for the remainder of the paper. 

If the RV acceleration is caused by a third body in the system, it could be detected in other data. We first measured the individual transit times from the \textit{TESS} dataset to search for \bedit{TTVs.} The resulting TTVs are shown in Figure~\ref{TTVs}. We do not identify any significant signals in the transit times, indicating that there is no evidence that the orbit of TOI-5800~b is being perturbed by another planet at the timescale of the \textit{TESS} observations.

We also investigated whether there is any evidence for an additional planet in the \textit{Gaia} astrometry. The \textit{Gaia} renormalized unit weight error (RUWE) describes the significance of astrometric excess noise in the astrometric solution, with RUWE~$>1.4$ being a conventional flag for significant astrometric variability \citep{Lindegren18}. In \textit{Gaia}~DR3, the RUWE for TOI-5800 is only 1.086, suggesting that there is no strong evidence for additional astrometric variability. To quantify the constraints from the \textit{Gaia} astrometric solution, we follow the methodology previously used in \citet{Limbach2024}. We adopt the limits on the astrometric reflex signal implied given an upper limit of RUWE~$<1.4$, and invert this to a limit on the astrometric perturbation following \citet{Belokurov2020, Korol2022}. We plot the resulting constraints in Figure~\ref{third_body_constraints}. In the region where the \textit{Gaia}~DR3 astrometry is most sensitive, between $2-3$~AU, we can rule out planets more massive than $\gtrsim$$4~M_J$. 

While the existence of an outer companion cannot be confirmed with the available data, we can quantify the 
potential third body parameters using the RV acceleration. We apply a simple sinusoidal model (equivalent to a zero-eccentricity Keplerian) with variable parameters of orbital period $P$, mass $M$, and inferior conjunction time $t_0$. We show the corresponding constraints on semi-major axis and $m\sin i$ in Figure~\ref{third_body_constraints}. If the hypothetical second planet has a minimal period similar to the observing baseline, it would only require $\sim$15~$M_\oplus$ to reproduce the acceleration; however, if the period is much larger than 1~year, then the $m\sin i$ may exceed $\gtrsim$1~$M_J$. Given that we observe a reversal in the RV acceleration, which only occurs twice per orbit, it is likely that the periodicity is not immensely greater than our observing baseline.

Though the quadratic trend is present in our RV dataset with high statistical significance, that does not necessarily indicate that it must have a planetary origin. In particular, linear and quadratic drifts in RVs can often be caused by stellar magnetic activity cycles. We investigated and did not observe a strong correlation between the quadratic radial velocity trend and the star's $S_{HK}$ stellar activity indicator, suggesting the quadratic trend may instead be due to a companion. Additionally, the observed trend occurs on timescales of hundreds of days, much longer than the star's expected rotation period ($\sim$20 days). However, we reserve judgment on the origin of the quadratic trend until a greater fraction of the RV signal can be observed. Confirming any outer companions in the system will therefore require additional follow-up observations.

\subsection{Mass, Radius, and Density of TOI-5800 b}
\begin{figure*}
\centering
\includegraphics[width=\textwidth]{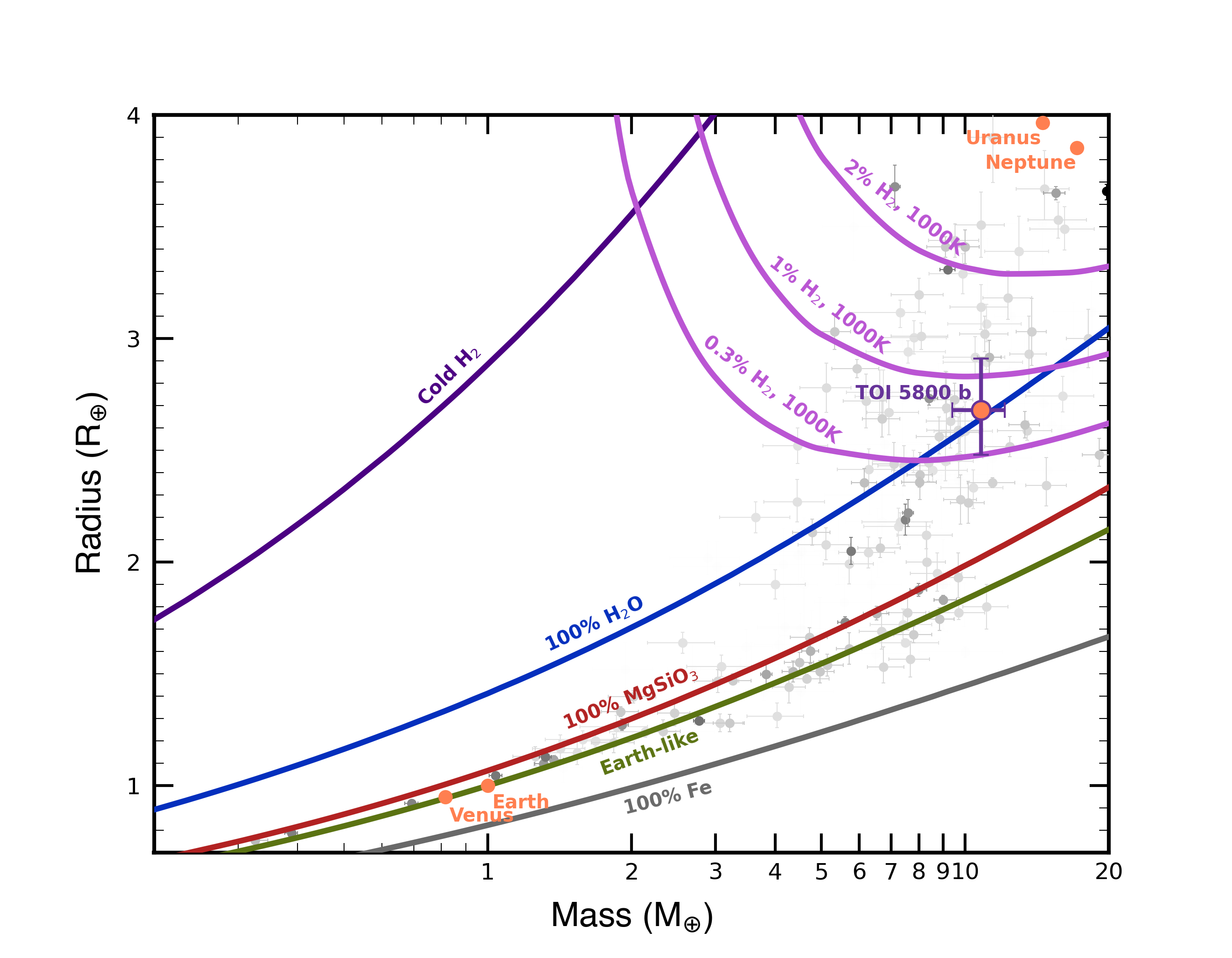}
\caption{Mass-radius diagram with known sub-Neptunes and super-Earths. We include planets from the NASA Exoplanet Archive (accessed April 3, 2025) that have masses constrained within 20\%. Theoretical models for a variety of interior compositions \citep{zen2016} and for a cold H$_2$ world \citep{sea2007} are shown as colored lines. Models for Earth-like cores with H$_2$ atmospheres are shown in light purple \citep{zen2018}. Solar system planets are shown in orange. \label{MR}}
\end{figure*}

Our PFS radial velocity observations allow us to measure the mass of TOI-5800 b and constrain its bulk composition for the first time. Our modeling reveals that TOI-5800 b has a mass and radius of \bedit{10.8$^{+1.3}_{-1.4}$ M$_{\oplus}$ and 2.68$^{+0.23}_{-0.20}$ R$_{\oplus}$,} respectively. This corresponds to a bulk density of \bedit{3.16$^{+0.86}_{-0.73}$} g cm$^{-3}$. We compare TOI-5800 b to other planets with well-constrained masses and to several model M-R curves in Figure~\ref{MR}. We include models for 1000 K Earth-like worlds with H$_2$ atmospheres from \citet{zen2018}. This model temperature is close to our estimated equilibrium temperature of 1119$^{+9}_{-8}$ K, which is calculated assuming an albedo of zero. \bedit{We find that TOI-5800 b's mass and radius are consistent with either a $<$1\% H$_2$ atmosphere with an Earth-like rocky core \citep{zen2018} or a composition including heavy volatiles such as water, methane, and ammonia.} However, these models assume no contribution from tidal heating. We \bedit{explore possible consequences of tidal heating in \S\ref{sec: tidal heating} and summarize all constraints on the planet's composition in \S\ref{Composition}.}

\subsection{Eccentricity of TOI-5800 b}
We find that TOI-5800 b has an unexpectedly high eccentricity of \bedit{0.39$\pm$0.07} and that our observations rule out circular orbits at more than $5\sigma$ confidence. The high eccentricity is consistent across both of our MCMC analyses. We find that the detection of non-zero eccentricity is largely driven by the RVs, but that the transit timing and duration is also consistent with this eccentricity. This high eccentricity is surprising given the planet's short orbital period. \bedit{We propose two explanations for TOI-5800 b's unusual dynamics: either it is being actively perturbed by an outer companion or it is undergoing HEM. 
 We investigate whether an outer planet could feasibly induce such high eccentricities in \S\ref{Companion_theory}. If TOI-5800 b is instead undergoing HEM, then it may provide insight into how sub-Neptunes migrate into the Neptune desert. We discuss consequences for the wider sub-Neptune population in \S\ref{History}.}



\section{Dynamical Analysis}
\label{State}

TOI-5800 b has an unusually high eccentricity compared to other Neptunian planets. It is therefore important to investigate how this eccentricity could have been excited, and how it affects the planet's physical properties. \bedit{We first investigate how the planet's orbit will eventually circularize due to tidal effects in \S\ref{Circularization0}. We then explore whether an outer companion could be driving TOI-5800 b's high eccentricity in \S\ref{Companion_theory}. To better understand the planet's physical properties, we then discuss radius inflation due to tidal heating in \S\ref{sec: tidal heating} and mass loss due to photoevaporation in \S\ref{Photoevaporation}.}

\subsection{Tidal Circularization}
\label{Circularization0}
TOI-5800 b's high eccentricity indicates that it is actively undergoing tidal circularization. We calculate its tidal circularization timescale $t_c$ from $M_{\star}$, $M_P$, $R_P$, and $a$ \citep{gold1966, ras1996}:
\begin{equation}
t_{c}=\frac{4Q'_p}{63}\frac{M_P a^{13/2}}{(GM_\star^3)^{1/2}R_P^5}
\end{equation}
 $Q'_p$ is the reduced tidal quality factor, defined as $Q'_p = 3 Q_p/2k_2$ where $Q_p$ and $k_2$ are the tidal quality factor and Love number, respectively. Assuming a $Q'_p$ value of 10$^5$, consistent with the values for Neptune and Uranus \citep{tit1990, ban1992}, we estimate $t_c$  to be $\sim$1 Gyr.  This suggests that it either migrated inward to its current orbit within the last $\sim$Gyr, or that it is actively being perturbed by an outer companion. As discussed in \S\ref{Companion}, we find tentative evidence for a planetary companion. However, this outer planet remains unconfirmed. We explore the possibility of an outer perturber in more detail in \S\ref{Companion_theory}. 
 
\begin{figure*}
\centering
\includegraphics[width=.95\textwidth]{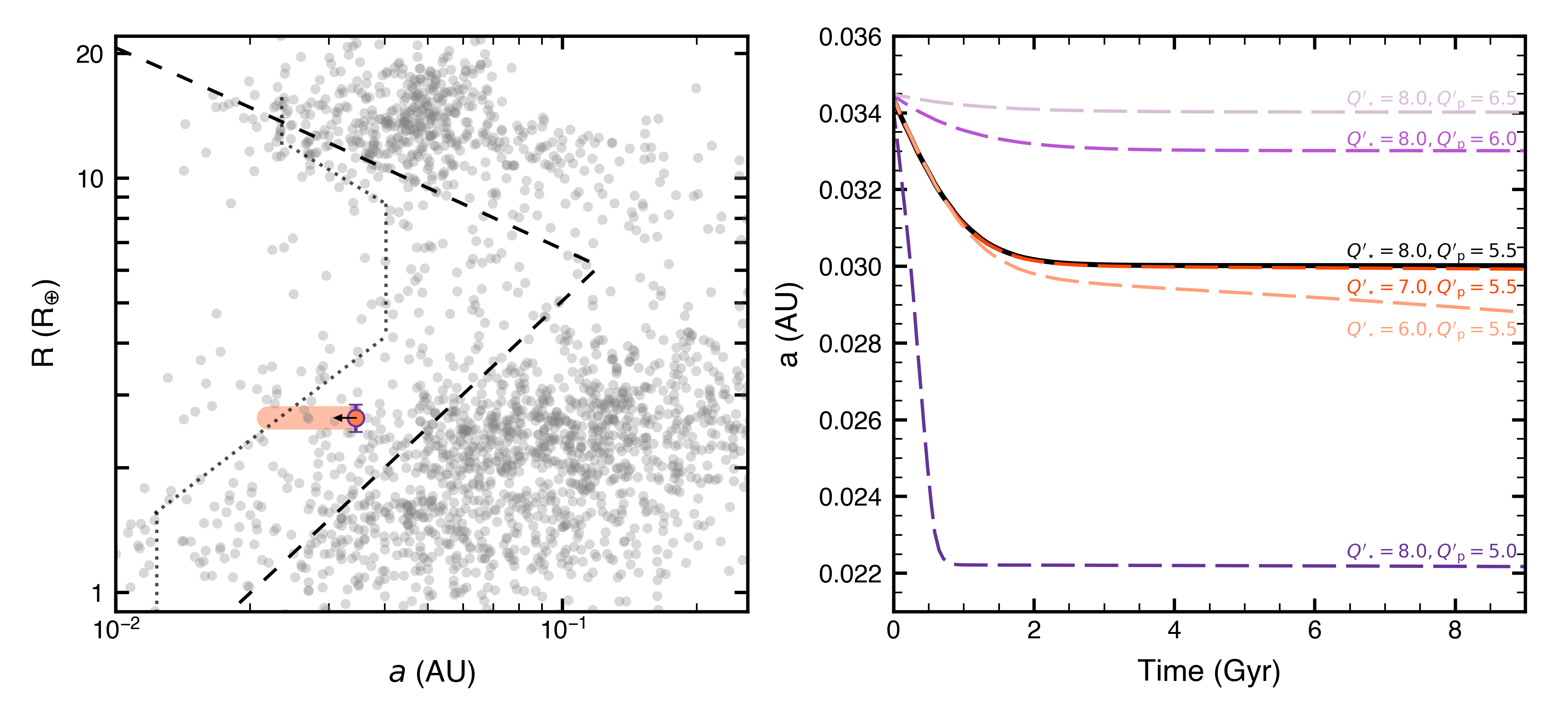}
\caption{Left: Radius and semi-major axis of planets listed in the NASA Exoplanet Archive. The \bedit{Neptune} desert is demarked by the black dashed lines \citep{maz2016}. \bedit{We also include updated boundaries from \citet{cas2024}, denoted by the black dotted lines.} TOI-5800 b is shown in orange. We expect the planet to circularize to a closer-in orbit, indicated by the black arrow. Right: Semi-major axis of TOI-5800 b as a function of time. Models for different values of $Q_{\star}$ and $Q_{\mathrm{p}}$ are shown in orange and purple, respectively. Assuming $Q_{\star} = 8.0$ and $Q_{\mathrm{p}} = 5.5$, we estimate a final semi-major axis of $\sim$0.03 AU. \label{Circularization}}
\end{figure*}

 We also predict a range of final orbital separations using MIST stellar evolutionary tracks to model the evolution of the host star. We evolve the planet’s semi-major axis and eccentricity with a 4th-order Runge-Kutta integration method that includes the \citet{gold1966} tidal prescription with static $Q'_{\star}$ and $Q'_{\mathrm{p}}$ parameters. TOI-5800 b's predicted orbital evolution is shown in Figure~\ref{Circularization}. Depending on the chosen values of $Q'_{\star}$ and $Q'_{\mathrm{p}}$, we estimate that the planet could eventually circularize to a semimajor axis as close in as $\sim$0.022 AU. TOI-5800 b is therefore an example of a planet actively migrating deeper into the Neptune desert. 


\subsection{Can an Outer Companion Explain TOI-5800 b's High Eccentricity?}
\label{Companion_theory}
To investigate whether an outer companion could be responsible for TOI-5800 b's high eccentricity, we first use Laplace-Lagrange secular theory to quantify perturbations induced by a hypothetical outer planet in the absence of tides (\S\ref{subsec:LL sol}). We then include tidal evolution (\S\ref{sec: sec perturbations + tides}). Finally, we examine whether an undetected outer companion within the constraints of our RV measurements could plausibly explain TOI-5800 b’s high eccentricity.

\subsubsection{Secular Dynamics with the Laplace-Lagrange Solution} \label{subsec:LL sol}
Secular dynamics can be used to study the long-term gravitational interactions between planets by averaging over short-term orbital motions. For small eccentricities and inclinations, the secular equations can be linearized. This leads to the Laplace-Lagrange solution, which predicts periodic oscillations in the eccentricity and inclination with a fixed amplitude \citep{1999ssd..book.....M}. 
We note that a planet consistent with the quadratic trend described in \S\ref{Companion} would likely be too far away to excite TOI-5800 b to its current eccentricity. To determine whether TOI-5800 b's eccentricity can instead be explained by a closer-in perturbing companion, we model an outer planet with a mass and period of 30 M$_\oplus$ and 60 days, respectively. These values correspond to an RV semi-amplitude of $\sim$5.7 m s$^{-1}$, making it a conservative example of a massive companion that might be undetected in our RV data. 
We set the stellar and inner planet parameters to the values listed in Tables \ref{tab:Stellar_properties} and \ref{tab:Planet_properties}, and we set the difference in the longitude of periapse between the two planets to $\Delta \omega = 180^\circ$, as this configuration leads to the highest mean eccentricity of the inner planet. We use this as an edge case for further dynamical analysis. 


\subsubsection{Secular Perturbations with Tidal Evolution} \label{sec: sec perturbations + tides}

The calculations performed thus far are incomplete because they have not yet included tidal perturbations. We now improve our secular model by incorporating tidal evolution. 
For a planet with an eccentric orbit, differential gravitational forces induce time-varying distortions in the planet's shape \citep{1981A&A....99..126H}.  Typically, tidal dissipation caused by planetary tides decreases both the orbital period and eccentricity of the planet \citep{1963MNRAS.126..257G, 2010A&A...516A..64L}. 
The eccentricity damping rate is approximated as:
\begin{equation} \label{ecc damping}
    \lambda_j = -\frac{\dot{e}_j}{e_j} = \frac{63}{4}\frac{1}{Q'_p}\frac{M_\star}{M_{p,j}}\left(\frac{R_{p,j}}{a_j}\right)^5n_j
\end{equation}
where $j = 1, 2$ corresponds to parameters defined for the inner and the outer planet, and $n_j$ is the mean motion \citep{1999ssd..book.....M, 2013ApJ...778....6Z}. For this analysis, we set $Q'_p = 10^5$, as in \S\ref{Circularization0}.

Because circularization timescales ($\sim 10^9$ yrs) are generally much longer than secular timescales ($\sim 10^3$ yrs for this system), tidal damping can be treated as a perturbation to the secular solution. We adopt the mathematical framework from \citet{2013ApJ...778....6Z}. Specifically, \citet{1999ssd..book.....M} define a coefficient matrix \textbf{A} that depends on $M_\star$, $M_{p,1}, M_{p,2}$, $a_1$, and $a_2$. Its eigenvalues directly determine the frequencies of the eccentricity oscillations under secular perturbations. To incorporate eccentricity damping (Equation \ref{ecc damping}), \citet{2013ApJ...778....6Z} introduce an additional imaginary term to the diagonal elements of the \textbf{A} matrix, representing the constant eccentricity damping rate. Therefore, the diagonal elements of the \textbf{A} matrix are modified as
\begin{equation}
    A_{jj} \rightarrow A_{jj} + i\lambda_j.
\end{equation}
The imaginary components of the eigenvectors of the modified \textbf{A} matrix are generally very small ($\lesssim 10^{-5}$). Therefore, for simplification, we ignore the imaginary part of the eigenvectors in our analysis. Hereafter, this allows us to follow the same steps as in the Laplace-Lagrange solution defined by \citet{1999ssd..book.....M}, with the addition of the aforementioned modifications. 

\begin{figure}
    \centering
    \includegraphics[width=0.48\textwidth]{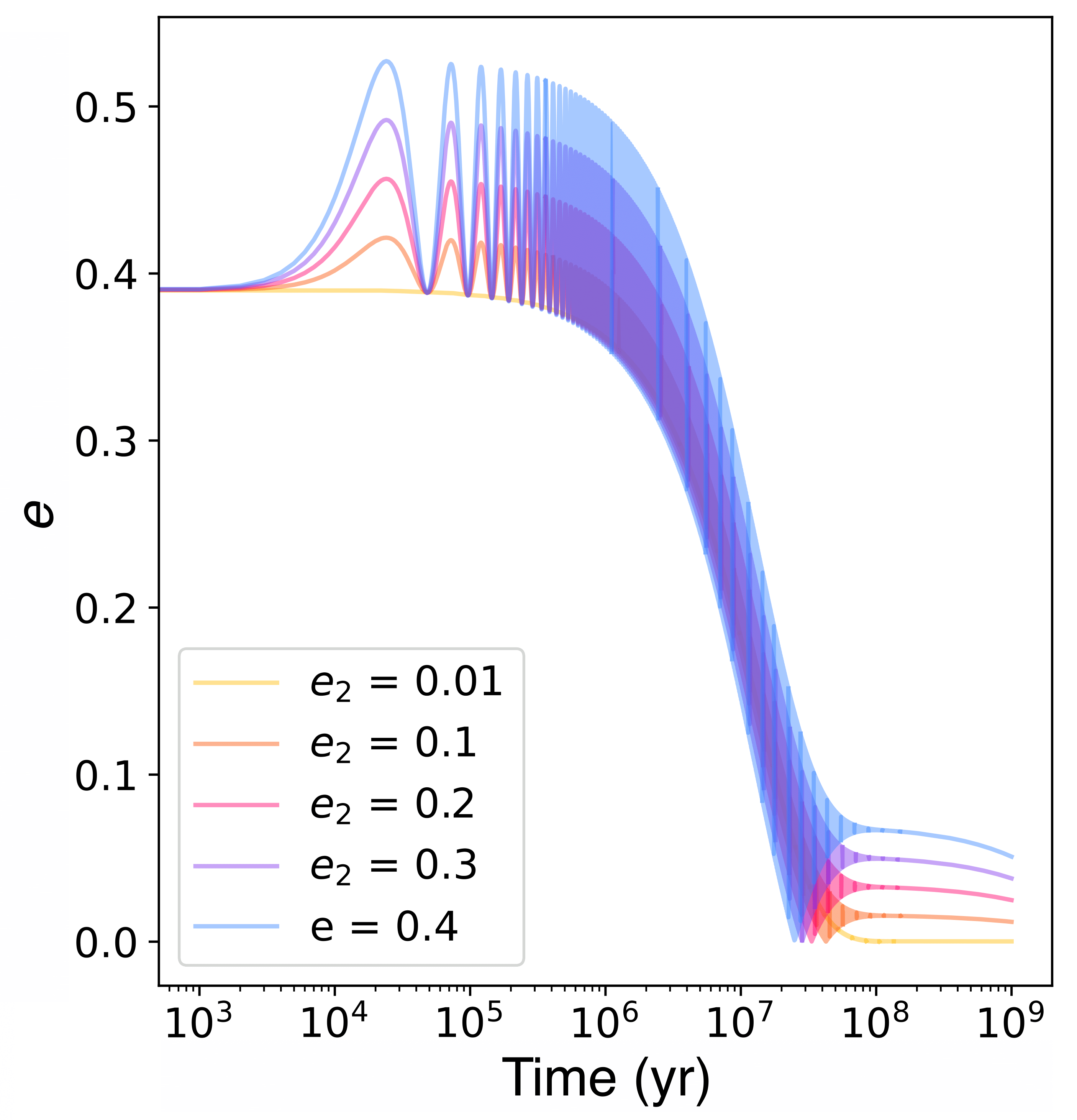}
    \caption{Eccentricity evolution of TOI-5800 b, assuming a hypothetical outer companion ($M_{p,2} = 30 M_\oplus$, $P_2 = 60$ days). 
    We show evolution curves for different outer planet eccentricities using the secular perturbation with tidal evolution model. We find that, regardless of the outer companion's eccentricity, TOI-5800 b always stabilizes to a relatively low ($<$0.1) eccentricity.}
    \label{fig:ecc_time}
\end{figure}

We run the secular perturbations and tidal evolution model for our two-planet system, considering a range of different outer planet eccentricities ($e_2 = 0.01, 0.1, 0.2, 0.3$ and $0.4$). We do not explore eccentricities higher than 0.4, as the assumptions of the Laplace-Lagrange framework are valid only for small eccentricities. 
The radius of the outer planet is determined using the mass-radius relationships given in \citet{2024A&A...686A.296M}. We initialize our model with the present-day orbital parameters of the inner planet (i.e., its observed semi-major axis and eccentricity) and evolve the system forward for 1 Gyr. The results are shown in Figure~\ref{fig:ecc_time}.

We find that the inner planet should reach a stable eccentricity within $\lesssim100$ Myr. The final eccentricity ranges between $0$ and $\sim0.06$, depending on the eccentricity of the outer companion. While the inner planet can retain a non-zero eccentricity due to an outer companion, we find that an eccentricity as high as the present-day value cannot be retained over long timescales for the tested outer planet masses and periods. This is true even in the most extreme case, where the outer companion has $e_2 = 0.4$. Therefore, we conclude that an outer companion alone cannot maintain the high eccentricity of TOI-5800 b. This suggests that additional dynamical processes
may be at play. Alternatively, we may be observing the planet at a special time, during a phase of rapid eccentricity damping.

There are several caveats to our model. Because our approximations are only valid for low eccentricities and inclinations, introducing correction terms would improve the model's accuracy. Larger deviations in the mutual inclination and the outer companion's eccentricity could potentially lead to a higher inner planet eccentricity. However, it is still unlikely that these effects alone could explain the observed eccentricity of TOI-5800 b. 


In addition to secular oscillations in eccentricity, another mechanism that can maintain non-zero orbital eccentricities is mean motion resonance (MMR). Mean motion resonance can maintain planetary eccentricities even in systems with tidal damping \citep{Luger2017}, as the forced eccentricity is maintained by resonant interactions even as inner planets lose eccentricity due to tidal interactions. While MMR is generally seen in the exoplanet population with low-to-moderate eccentricities \citep{Lithwick2012}, a parameter space of stable MMRs also exists at high ($e\sim0.3-0.4$) orbital eccentricities \citep{Chiang2003, Malhotra2016, Wang2017, Tamayo2024}.
However, resonant configurations generally produce significant transit timing variations \citep[e.g.,][]{Mills2016, Luger2017,Agol2021, Weisserman2023, MacDonald2023, Goldberg2023}. While only a limited baseline of data is available for TOI-5800 b, we do not observe any significant TTVs (see Figure~\ref{TTVs}), which is evidence against TOI-5800 b being in MMR with an exterior companion planet. Detailed MMR calculations are beyond the scope of this work. 

\subsection{Radius Inflation due to Tidal Heating} \label{sec: tidal heating}
TOI-5800 b's eccentricity affects not only its orbital evolution, but also its physical properties. 
Tidal heating can drive atmospheric inflation, a phenomenon previously observed on other sub-Neptunes \citep{2019ApJ...886...72M, mil2020, 2022ApJ...931L..15S}. This inflation can also enhance atmospheric mass loss through photoevaporation \citep{2018MNRAS.479.5012O, 2021A&A...647A..40A}. 

To investigate how tidal heating impacts TOI-5800 b, we predict the amount of radius inflation it may be experiencing at the present day. We use a tidal model based on the viscous approach to equilibrium tide theory \citep{2010A&A...516A..64L} to estimate TOI-5800 b's tidal luminosity $L_{\rm tide}$, defined as the rate of energy dissipation through tidal heating. Assuming zero planetary obliquity and $Q_p' =10^5$, we find \bedit{$L_{\rm tide} \approx 3.4 \times 10^{26}$ erg s$^{-1}$, which corresponds to about 10\% of the incident stellar power.} 

We quantitatively assess the radius inflation resulting from this tidal luminosity using empirical relations described by \citet{2019ApJ...886...72M}. These relations provide an approximate estimate of the degree of radius inflation based on fits to planetary thermal structure simulations for a planet with a rocky core and a H/He envelope. 
\bedit{Using these empirical relations, we find that with an envelope mass fraction $f_{\rm env}$ of just $\sim 0.033\%$, TOI-5800 b expands to $\sim4.4$ times larger than it would be in the absence of tidal heating.} This corresponds to the planet's Hill radius, meaning that 
atmospheric expansion beyond this limit would render the planet unable to retain its atmosphere. If the planet's atmosphere is composed of H/He, we therefore expect tidal heating to drive substantial atmospheric expansion even for extremely small envelope mass fractions. \bedit{As a result, $f_{\rm env}$ is constrained to be $\lesssim$0.033\%.} 
We note that using these empirical relations is necessary because a direct fit to the simulation results does not converge, likely because the atmosphere is not stable to photoevaporation (as discussed in \S\ref{Photoevaporation}). These results should therefore be considered with caution. 

\subsection{Photoevaporation}
\label{Photoevaporation}
Close-in planets like TOI-5800 b experience intense irradiation from high-energy UV and X-ray photons emitted by their host stars. Here we predict the present-day atmospheric mass loss rate from photoevaporation. We first estimate the star's X-ray luminosity $L_X$ from the measured $R^\prime_{HK}$ value (see \S\ref{Stellar Age}). Using the relation for K dwarfs from \citet{hou2017}, we find $L_X\sim1.9\times10^{28}$ erg s$^{-1}$. 
\bedit{We then calculate the mass loss using the  Planetary Atmospheres and Stellar RoTation RAtes \citep[{\tt PASTA},][]{bon2021} code.} {\tt PASTA} simulates atmospheric evolution using hydrodynamic modelling as the basis for estimating mass loss rates. Specifically for this work, we interpolated across the large grid of upper atmosphere hydrodynamic models of \citet{kub2018} and \citet{kub2021} using the machine learning interpolation algorithm of \citet{rez2025}. We find that equilibrium temperature, XUV flux and mass loss rate vary significantly over the course of the eccentric orbit, and that the mass loss rate increases with increasing XUV flux and tidal heating, as expected. Assuming an albedo of zero, we estimate mass loss rates in the range $(0.7 - 4.0)\times10^{10}$ g s$^{-1}$ between apastron and periastron. If we include a tidal heating contribution of 100 K, the mass loss rate can increase up to $4.4\times10^{10}$ g s$^{-1}$. 
\bedit{For the maximum possible envelope mass fraction $f_{\rm env} \approx 0.033\%$ (as derived in \S\ref{sec: tidal heating}), we estimate an envelope mass $M_{\rm env} = M_pf_{\rm env} \approx 0.0036$ $\mathrm{M}_\oplus$. Given the predicted mass loss rate, we expect such an envelope to be lost on 10 Myr timescales.} Therefore, unless TOI-5800 b migrated into Neptune Desert within the last \bedit{few} 10 Myrs, we expect the planet to have lost most of its primordial H-dominated atmosphere. This prediction is also strengthened by the fact that its host star would have had an even higher XUV luminosity when it was young, and that the planet likely retains residual heat from its formation. 

We note that \bedit{this calculation} of the mass loss rate assume a H/He-dominated envelope, making them only approximate. If the planet instead has an atmosphere with heavy volatiles, such as a steam-dominated atmosphere, the mass loss rate would be lower \citep[e.g.,][]{egg2024,egg2025}. 
A steam atmosphere could also cool more efficiently \citep[e.g.,][]{joh2020, yos2022, gar2023}, reducing atmospheric expansion. This would lead to a higher envelope mass fraction, as predicted by the planetary interior model accounting for tidal heating. As a result, the planet would retain a more massive envelope, increasing the timescale for complete atmospheric loss compared to a H/He atmosphere. Therefore, rigorous modeling, accounting for atmospheric composition and the degree of tidal heating, is required for a more accurate estimate of atmospheric mass loss.

\section{Discussion}
\label{Discussion}

\begin{figure*}
\centering
\includegraphics[width=.95\textwidth]{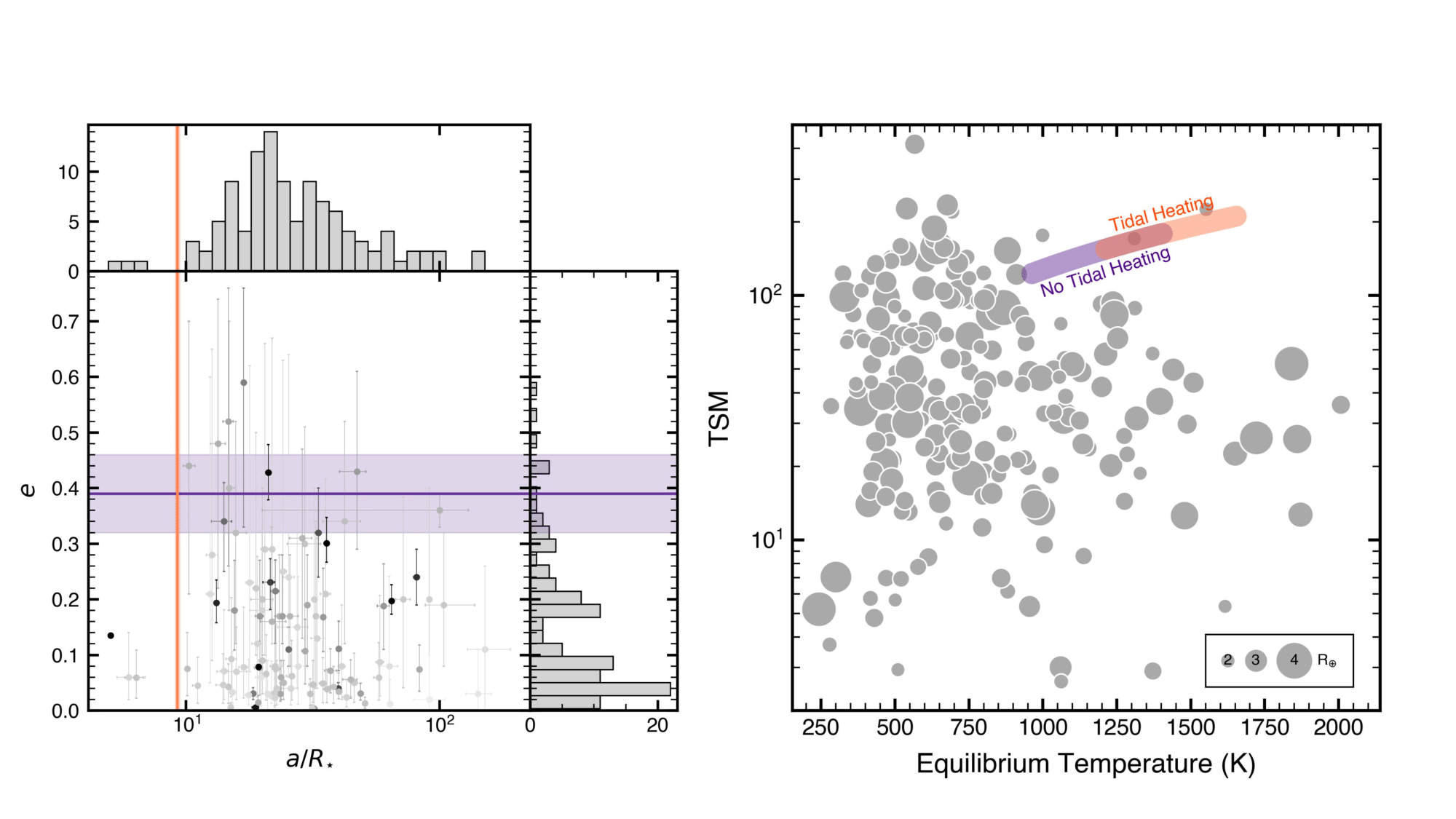}
\caption{Left: Eccentricity and semi-major axis  in units of stellar radii for all confirmed planets between 1.9-4.0 R$_{\oplus}$ listed on the NASA Exoplanet Archive. TOI-5800 b $e$ and $a/R_{\star}$ values are indicated by purple and orange lines, respectively. 1$\sigma$ uncertainties are indicated by the shaded regions. Right: TSM values and effective temperatures for all confirmed planets between 1.9-4.0 R$_{\oplus}$ with available measurements. Because of its eccentric orbit, the planet may experience a range of temperatures. Using revised values for the radius and mass of the planet, we show the range of equilibrium temperatures and TSM values between apocenter and pericenter in purple. Additionally, the planet is likely experiencing tidal heating. We show the resulting range in orange, assuming $Q'_p=10^4$ \citep{mil2020}. \label{Discussion_fig}}
\end{figure*}

\subsection{Eccentric Neptune Population} 
\label{History}
\bedit{Small ($\lesssim$3.5 R$_{\oplus}$) planets tend to have lower eccentricities, on average, compared to larger planets \citep[0.05$\pm$0.01 compared to 0.20$\pm$0.03, respectively, ][]{gil2025}. However, a population of close-in 
Neptunian planets have been found on eccentric or polar orbits \citep[e.g.,][]{cor2020, bou2023}. In particular, systems with only one known transiting planet tend to have higher eccentricities than systems with multiple transiting planets \citep{van2019}. } 

Two mechanisms that may be responsible for sculpting \bedit{the eccentric Neptune population} are HEM and atmospheric evaporation. \bedit{Because} most close-in sub-Neptunes have eccentricities consistent with zero, \bedit{it is} difficult to probe how HEM shapes the Neptune desert. With an eccentricity $>5 \sigma$ above zero, TOI-5800 b is comparatively extreme. \bedit{This is demonstrated in Figure~\ref{Discussion_fig}, where we compare TOI-5800 b to the wider population of sub-Neptunes. TOI-5800 b} may therefore provide direct insight into how close-in Neptunes migrate inwards, making it critical to understand its dynamical history.
 Though we find tentative evidence for an additional planet, any outer companion is unlikely to cause the observed perturbations in TOI-5800 b's orbit (see \S\ref{Companion_theory}). We therefore propose that TOI-5800 b is actively undergoing HEM into the Neptune desert. 
 If TOI-5800 b is undergoing HEM, then it provides direct evidence that HEM plays a key role in shaping the Neptune desert. In addition, it makes TOI-5800 b an important planet for future follow-up work probing the dynamics of the desert and understanding the impact of photoevaporation on migrating planets.

\subsection{Composition of TOI-5800 b}
\label{Composition}
The internal and atmospheric compositions of sub-Neptune planets remain unclear. They may be predominantly rocky, be dominated by heavy volatile elements, have large gaseous envelopes, or host planet-wide oceans. As shown in Figure~\ref{MR}, TOI-5800 b's mass and radius are consistent with either a $<$1\% H$_2$ atmosphere with an Earth-like rocky core or a composition including heavy volatiles. However, as described in \S\ref{sec: tidal heating}, we expect even a small amount of H/He in the atmosphere to result in significant inflation, placing an upper constraint on the H/He envelope mass fraction (\bedit{$f_{env}<$0.033\%}). We therefore predict that, if TOI-5800 b does have a H/He atmosphere, it must have migrated inward relatively recently, before a significant amount of H/He could be tidally heated and lost. We propose that a more likely explanation is that TOI-5800 b's atmosphere is dominated by heavy volatiles such as water, methane, and ammonia. In this scenario, we expect less radius inflation due to tidal heating, and therefore less mass loss.


\subsection{Potential for Atmospheric Characterization} 
\label{JWST}

Follow-up atmospheric characterization could answer many of the outstanding questions about this system. For instance, constraints on TOI-5800 b's atmospheric composition and mass loss rate would provide key insight into the planet's dynamical history. If an outer companion scenario is ruled out, then characterizing this system will also help probe the mechanisms that sculpt the Neptune desert. 

The \textit{James Webb Space Telescope} (\textit{JWST}) provides an unparalleled opportunity to study the atmospheres of sub-Neptune planets with transmission spectroscopy. Because of its large aperture and wavelength coverage into the mid-infrared, it is sensitive to molecules such as water, methane, carbon dioxide, and nitrogen dioxide. 
TOI-5800 b is a prime target for transmission spectroscopy with \textit{JWST}, with a Transmission Spectroscopy Metric (TSM) of 164. This metric, first proposed by \citet{kem2018}, is used to describe how well suited a planet is to atmopsheric characterization, with higher metric values corresponding to more suitable planets. \citet{hor2023} used the TSM and Emission Spectroscopy Metric (ESM) to systematically identify \textit{TESS} planets ideal for \textit{JWST} follow-up. They found that, within its equilibrium temperature and radius regime ($T_{eq} = 800 - 1250$ K, $R_P = 2.75 - 4$ R$_\oplus$), TOI-5800 b is the best candidate for transmission spectroscopy observations. None of the other top five planet candidates in TOI-5800 b's temperature and radius regime have been observed with \textit{JWST}, and only one (TOI-1339 b) has a constrained mass \citep{bad2020, lub2022, ore2023, pol2024}. In addition, one other top-ranked candidate (TOI-4337 b) has since been designated a false positive \footnote{\url{https://exofop.ipac.caltech.edu/tess/target.php?id=399642071}, \url{https://exofop.ipac.caltech.edu/tess/edit_obsnotes.php?id=399642071}}. 

Given TOI-5800 b's eccentric orbit, its equilibrium temperature during transit events may differ from its previously reported value of 1119 K. We estimate the equilibrium temperatures at apastron and periastron to be $\sim$1000 and 1400 K, respectively, corresponding to TSMs of 156 and 111. This range of values is shown in Figure~\ref{Discussion_fig}. 
Additional tidal heating may make TOI-5800 b an even better candidate for transmission spectroscopy. Depending on the assumed value of $Q'_p$, we expect up to several hundred degrees of heating. We demonstrate how this could affect the resulting TSM values in Figure~\ref{Discussion_fig}. 

\citet{hor2023} also identified TOI-5800 b as the best candidate for emission spectroscopy observations within its temperature and radius regime, measuring an ESM of 21. We estimate an occultation impact parameter $b_{occ}$ of 0.51$^{+0.12}_{-0.09}$, indicating that, despite its eccentricity, we expect TOI-5800 b's secondary eclipse to be observable. Given our constraints on the planet's orbital parameters, we predict that the secondary eclipse occurs 21.4$\pm$1.1 hours before transit. Secondary eclipse observations would simultaneously improve our estimate of TOI-5800 b's eccentricity while providing a constraint on the temperature of the planet. Because tidal heating contributions may be significant, this would also provide a unique opportunity to test tidal heating models. 

Additionally, phase curve measurements would provide insight into TOI-5800 b's heating and cooling processes. Between apastron and periastron, the amount of flux received increases by a factor of $((1+e)/(1-e))^2=5.2$. We therefore expect the planet's temperature, chemistry, and emission spectrum to change throughout the orbit.  Transient heating has previously been measured for eccentric gas giants, such as HD 80606b \citep{lau2009, wit2016} and HAT-P-2b \citep{lew2013, wit2017}. TOI 5800-b's phase curve would allow us to extend these findings to the sub-Neptune regime. Such observations would also improve the constraints on the planet's eccentricity and argument of periastron. 

Follow-up observations could also search for signatures of atmospheric escape. Extended atmospheres of neutral hydrogen and metastable helium have previously been detected on other warm Neptunes \citep[e.g.,][]{2015Natur.522..459E, 2018A&A...620A.147B, 2022NatAs...6..141B, 2022AJ....164..234V}, and are detectable in the ultraviolet and infrared regimes, respectively. Given the TOI-5800 system is only $\approx$42.9 pc away and has a \textit{Gaia} systematic RV of $-$31.7 km s$^{-1}$, Lyman-$\alpha$ transmission spectroscopy should be possible with the \textit{Hubble Space Telescope} ($HST$). If future observations identify an escaping envelope around TOI-5800, it would provide further evidence that atmospheric evaporation plays a critical role in shaping the evolution of warm Neptunes.



\section{Conclusion}
\label{Conclusion}
In this paper, we report the confirmation of TOI-5800 b, a planet entering the Neptune desert. The main conclusions of \bedit{the} paper are summarized as follows: 

\begin{enumerate}
\item[\textbullet] TOI-5800 b is a confirmed sub-Neptune planet with a mass of 10.8$^{+1.3}_{-1.4}$ M$_{\oplus}$ and radius of 2.68$^{+0.23}_{-0.20}$ R$_{\oplus}$. 
\item[\textbullet] TOI-5800 b is unusually eccentric given its close-in orbit. It has an eccentricity $>5\sigma$ above zero. 
\item[\textbullet] Due to its high eccentricity, TOI-5800 b is tidally migrating into the Neptune desert. We propose that TOI-5800 b is actively undergoing high eccentricity migration. However, additional work is needed to confirm whether there are any perturbing outer companions in the system. 
\item[\textbullet] TOI-5800 b is a top candidate for transmission and emission spectroscopy with $JWST$ and $HST$. Follow-up atmospheric characterization could probe its atmospheric composition, test tidal heating and transient heating models, and search for signatures of atmospheric escape. This would help us better constrain the dynamical history of the planet and provide insight into how planets enter the Neptune desert. 
\end{enumerate}

TOI-5800 b delivers on \textit{TESS}’s promise to uncover planets in previously sparse regions of parameter space. As a rare example of a planet migrating into the Neptune desert, it offers a unique opportunity to study the forces shaping planetary populations.

\section*{Acknowledgements}

This material is based upon work supported by the National Science Foundation Graduate Research Fellowship under Grant No. 1745302. The work of H.P.O., M.L., \& D.E. has been carried out within the framework of the NCCR PlanetS supported by the Swiss National Science Foundation under grants 51NF40\_182901 and 51NF40\_205606. D.E. and M.L. acknowledge funding support from the Swiss National Science Foundation projects 200021\_200726 (D.E.) and PCEFP2\_194576 (M.L.). C.B. acknowledges support from the Swiss Space Office through the ESA PRODEX program. T.G.W. acknowledges support from the UKSA and the University of Warwick. T.D. acknowledges support from the McDonnell Center for the Space Sciences at Washington University in St. Louis.

We acknowledge the use of public TESS data from pipelines at the TESS Science Office and at the TESS Science Processing Operations Center. Funding for the \textit{TESS} mission is provided by NASA's Science Mission Directorate. This research has made use of the Exoplanet Follow-up Observation Program website, which is operated by the California Institute of Technology, under contract with the National Aeronautics and Space Administration under the Exoplanet Exploration Program. This paper includes data collected by the \textit{TESS} mission that are publicly available from the Mikulski Archive for Space Telescopes (MAST). Resources supporting this work were provided by the NASA High-End Computing (HEC) Program through the NASA Advanced Supercomputing (NAS) Division at Ames Research Center for the production of the SPOC data products.

CHEOPS is an ESA mission in partnership with Switzerland with important contributions to the payload and the ground segment from Austria, Belgium, France, Germany, Hungary, Italy, Portugal, Spain, Sweden, and the United Kingdom. The CHEOPS Consortium would like to gratefully acknowledge the support received by all the agencies, offices, universities, and industries involved. Their flexibility and willingness to explore new approaches were essential to the success of this mission. CHEOPS data analysed in this article will be made available in the CHEOPS mission archive (\url{https://cheops.unige.ch/archive\_browser/}).

This research has made use of NASA's Astrophysics Data System Bibliographic Services, and the SIMBAD database, operated at CDS, Strasbourg, France. 

\section*{Data Availability}
\bedit{All the {\it TESS} data presented in this article were obtained from the Mikulski Archive for Space Telescopes (MAST) at the Space Telescope Science Institute. The specific observations analyzed can be accessed via \dataset[doi: 10.17909/06cf-qt93]{https://doi.org/10.17909/06cf-qt93}.}


\software{matplotlib \citep{plt}, 
          numpy \citep{np}, {edmcmc \citep{andrew_vanderburg_2021_5599854}}
          }



\vspace{5mm}
\bibliography{refs}{}
\bibliographystyle{aasjournal}

\clearpage
\appendix

\renewcommand{\thefigure}{A\arabic{figure}}
\renewcommand{\theHfigure}{A\arabic{figure}} 
\setcounter{figure}{0}
\renewcommand{\thetable}{A\arabic{table}}
\renewcommand{\theHtable}{A\arabic{table}}
\setcounter{table}{0}

\begin{figure*}
\centering
\includegraphics[width=.95\textwidth]{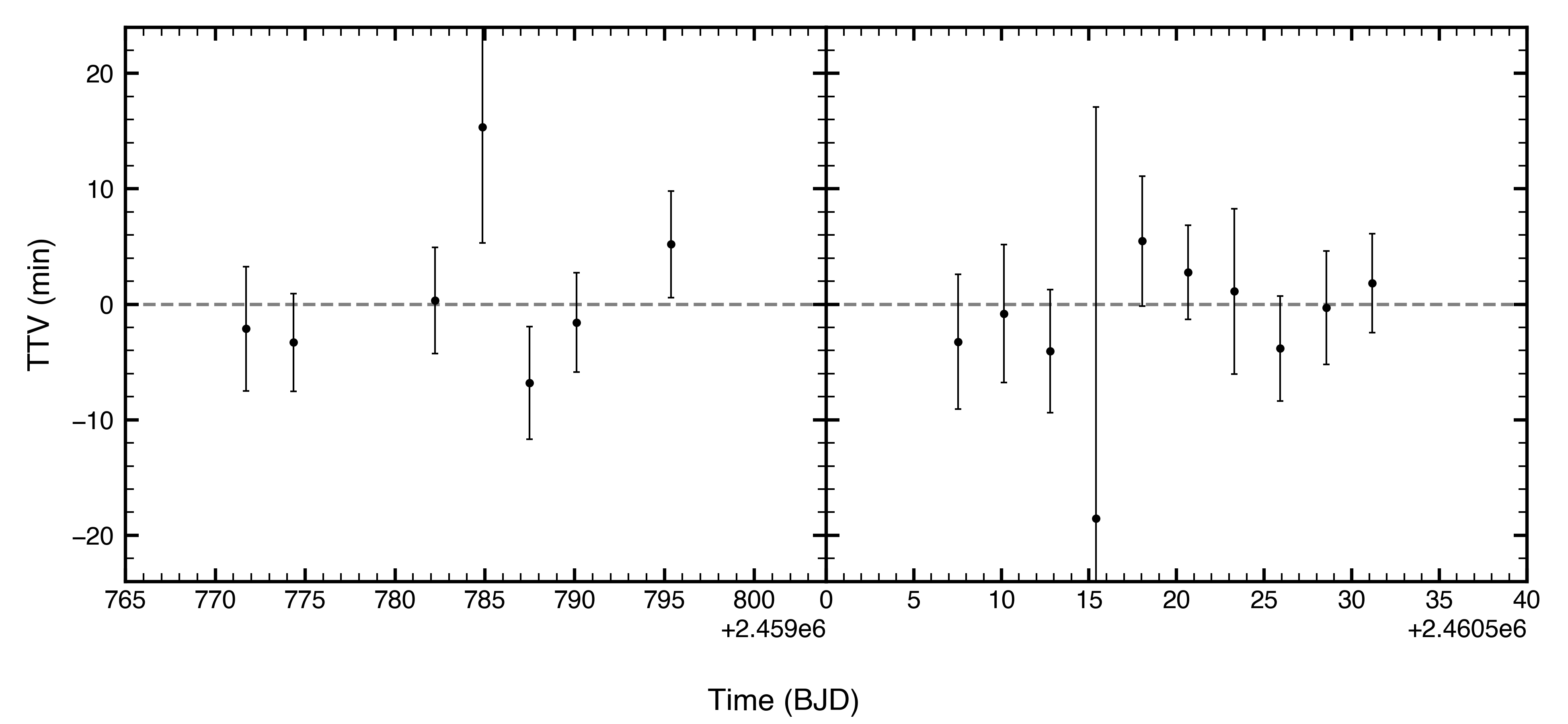}
\caption{Transit timing variations (TTVs) calculated for 17 transit events observed with \textit{TESS}. TTVs are measured relative to the linear ephemeris calculated using values listed in Table~\ref{tab:Planet_properties}. We do not observe any significant variations in the TTV data.   \label{TTVs}}
\end{figure*}

\begin{figure}
\centering
\includegraphics[width=\columnwidth]{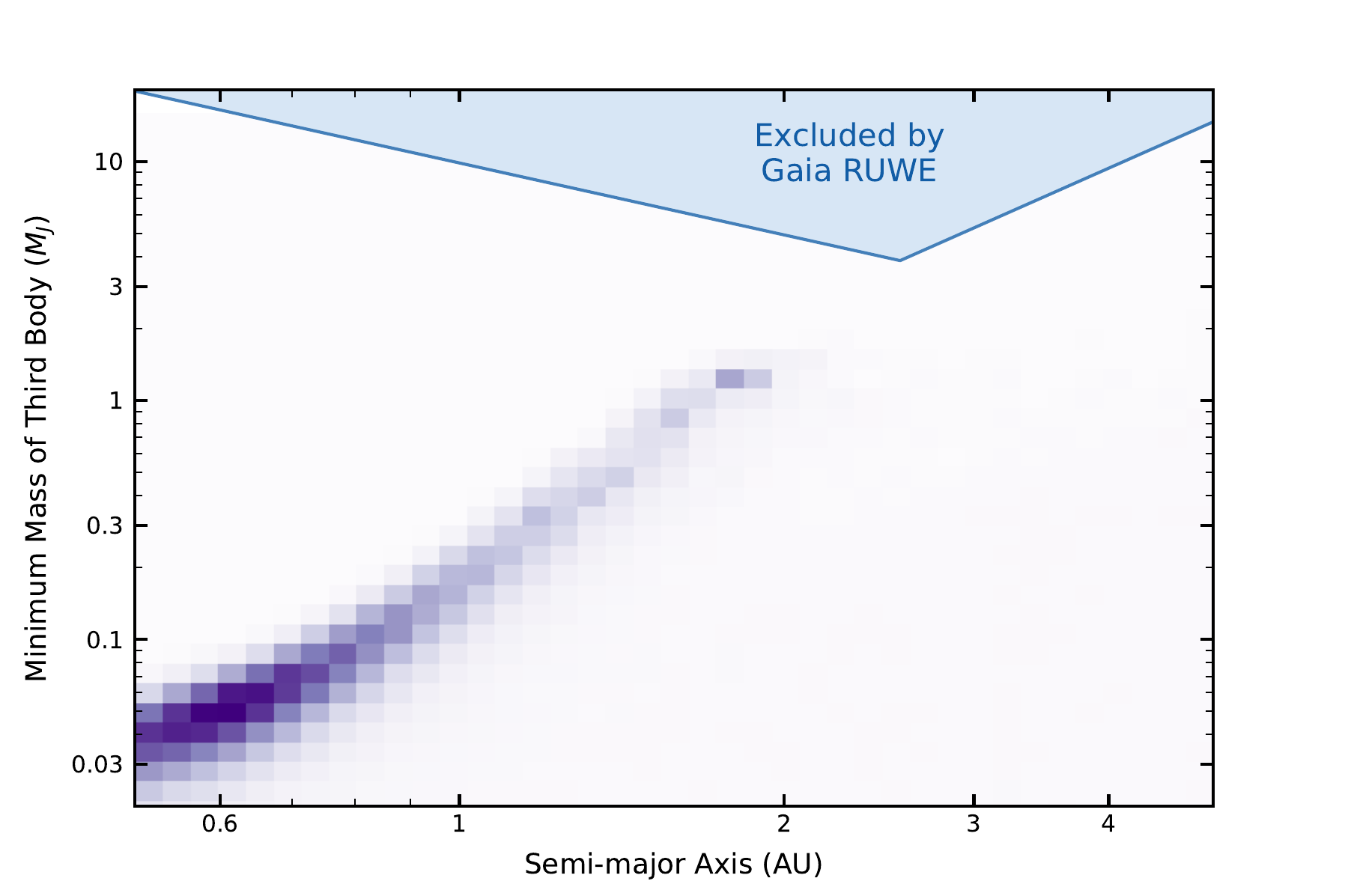}
\caption{Posterior distribution of potential outer companion's mass and orbital separation. The heatmap corresponds to constraints allowed by the RV acceleration under a simple zero-eccentricity model (described in \S\ref{Companion}). The region ruled out by the absence of a significant astrometric signal is highlighted in blue. Further observations are needed to determine whether there is an additional planet in this system.}
\label{third_body_constraints}
\end{figure}

\begin{table}
\centering
\caption{PFS radial velocity observations of TOI 5800.   } \label{tab:RVs}
\centering
\begin{tabular}{cc}
\hline
\hline
Time (BJD) & RV (m s$^{-1}$) \\
\hline

 2460456.81318 & 1.7 $\pm$ 0.9  \\ 
 2460459.82986 & 3.5 $\pm$ 0.8  \\ 
 2460460.86586 & -9.1 $\pm$ 0.8  \\ 
 2460461.84021 & -5.1 $\pm$ 0.8  \\ 
 2460462.86443 & -0.8 $\pm$ 0.8  \\ 
 2460463.82448 & -7.2 $\pm$ 0.7  \\ 
 2460488.76448 & 2.8 $\pm$ 0.8  \\ 
 2460490.75438 & -3.4 $\pm$ 0.7  \\ 
 2460491.71020 & 0.1 $\pm$ 0.7  \\ 
 2460492.74184 & -7.9 $\pm$ 0.7  \\ 
 2460509.74486 & 5.9 $\pm$ 0.7  \\ 
 2460515.74278 & 1.9 $\pm$ 0.9  \\ 
 2460515.75354 & 1.0 $\pm$ 0.9  \\ 
 2460516.69479 & 0.0 $\pm$ 0.8  \\ 
 2460536.64155 & -1.8 $\pm$ 0.7  \\ 
 2460537.72688 & -6.1 $\pm$ 0.8  \\ 
 2460545.63502 & -4.7 $\pm$ 1.0  \\ 
 2460546.63715 & 7.2 $\pm$ 1.2  \\ 
 2460546.64848 & 3.7 $\pm$ 1.2  \\ 
 2460598.51803 & -2.8 $\pm$ 0.8  \\ 
 2460599.54733 & 2.0 $\pm$ 0.7  \\ 
 2460604.51159 & 2.7 $\pm$ 0.8  \\ 
 2460604.52250 & 1.0 $\pm$ 0.8  \\ 
 2460605.53693 & -5.8 $\pm$ 0.8  \\ 
 2460605.54809 & -6.6 $\pm$ 0.9 \\ 
 
\hline
\end{tabular}
\end{table}
\begin{table}
\centering
\caption{TOI-5800 b Planet Properties.} \label{tab:Planet_properties}
\centering
\begin{tabular}{lllll}
\hline
Parameter & Units & Prior & \multicolumn{2}{c}{\bedit{Value}}   \\
& & &  1p Model & 2p Model  \\
\hline
\multicolumn{5}{c}{Fit Parameters} \\ 
\hline
$P$  & Period (days) & \bedit{$\mathcal{U}(2.622876, 2.632876)$} &2.627883$\pm$0.000005 & 2.627882$\pm$0.000005\\
$t_0$  & Time of inferior conjunction (BJD)& \bedit{$\mathcal{U}(2459771.7032, 2459771.7232)$} &2459771.7142$^{+0.0016}_{-0.0015}$ & 2459771.7147$^{+0.0014}_{-0.0014}$\\
$R$   & Radius ($R_{\oplus}$)& \bedit{$\mathcal{U}(0, 4.36)$} &2.58$^{+0.23}_{-0.18}$ & 2.68$^{+0.23}_{-0.20}$\\
$M$   & Mass ($M_{\oplus}$) & \bedit{$\mathcal{U}(0.0, 317.9)$} &9.6$^{+1.6}_{-1.6}$ & 10.8$^{+1.3}_{-1.4}$\\
$a/R_{\star}$ & Semi-major axis in stellar radii & \bedit{$\mathcal{U}(0, 20)$} &9.26$^{+0.14}_{-0.14}$ & 9.26$^{+0.14}_{-0.14}$\\
$\mathrm{cos}i$   & Cosine of inclination & \bedit{$\mathcal{U}(0, 1)$} &0.091$^{+0.010}_{-0.007}$ & 0.083$^{+0.006}_{-0.005}$\\
$\sqrt{e}$sin$\omega$ & & \bedit{$\mathcal{U}(-1, 1)$} &$-$0.31$^{+0.27}_{-0.19}$ & $-$0.48$^{+0.16}_{-0.10}$\\
$\sqrt{e}$cos$\omega$ & & \bedit{$\mathcal{U}(-1, 1)$} &0.48$^{+0.08}_{-0.08}$ & 0.39$^{+0.07}_{-0.06}$\\ 
$\mu_1$ & $TESS$-band linear limb-darkening coeff& \bedit{$\mathcal{N}(0.52, 0.20)$} & 0.60$^{+0.17}_{-0.18}$ & 0.57$^{+0.17}_{-0.17}$ \\
$\mu_2$  & $TESS$-band quadratic limb-darkening coeff& \bedit{$\mathcal{N}(0.10, 0.20)$} & 0.21$^{+0.18}_{-0.18}$& 0.17$^{+0.18}_{-0.18}$\\
$\mu_1$ & CHEOPS-band linear limb-darkening coeff& \bedit{$\mathcal{N}(0.65, 0.20)$} &0.69$^{+0.15}_{-0.16}$ & 0.66$^{+0.16}_{-0.17}$\\
$\mu_2$  & CHEOPS-band quadratic limb-darkening coeff& \bedit{$\mathcal{N}(0.06, 0.20)$} & 0.09$^{+0.17}_{-0.17}$ & 0.06$^{+0.17}_{-0.17}$ \\
$\sigma_{\mathrm{TESS}}$ & TESS jitter & \bedit{$\mathcal{U}(0, 0.01)$} &0.000015$\pm$0.000013 & 0.000015$\pm$0.000013 \\ 
$\sigma_{\mathrm{CHEOPS}}$ & CHEOPS jitter & \bedit{$\mathcal{U}(0, 0.01)$} &0.000277$\pm$0.000020 & 0.000277$\pm$0.000020 \\ 
$\sigma_{\mathrm{PFS}}$ & PFS jitter (m s$^{-1}$) & \bedit{$\mathcal{U}(0, 5)$} &2.47$\pm$0.45 & 1.91$\pm$0.40 \\ 
$\gamma_{\mathrm{TESS}}$ & TESS offset & \bedit{$\mathcal{U}(-1,1)$} &$-$0.000011$\pm$0.000016&$-$0.000011$\pm$0.000016 \\ 
$\gamma_{\mathrm{CHEOPS}}$ &CHEOPS offset & \bedit{$\mathcal{U}(-1, 1)$} &0.000113$\pm$0.000028 & 0.000111$\pm$0.000028 \\ 
$\gamma_{\mathrm{PFS}}$ & PFS offset (m s$^{-1}$)  & \bedit{$\mathcal{U}(-5, 5)$} &1.59$\pm$0.57 & $-$0.52$\pm$0.76\\ 
$\dot{\gamma}_{\mathrm{CHEOPS}}$ &CHEOPS linear term & \bedit{$\mathcal{U}(-0.1, 0.1)$} &$-$0.0009$\pm$0.0004& $-$0.0009$\pm$0.0004\\ 
$\dot{\gamma}_{\mathrm{PFS}}$ & RV linear term (m s$^{-1}$ day$^{-1}$) & \bedit{$\mathcal{U}(-1,1)$} &-- & 0.014$\pm$0.009\\
$\ddot{\gamma}_{\mathrm{PFS}}$ & RV quadratic term (m s$^{-1}$ day$^{-2}$) & \bedit{$\mathcal{U}(-1,1)$} &-- & $-$0.0007$\pm$0.0002\\
\hline
\multicolumn{5}{c}{Derived Parameters} \\ 
\hline
$a$  & Semi-major axis (AU) & & 0.0345$^{+0.0007}_{-0.0007}$ & 0.0345$^{+0.0007}_{-0.0007}$\\
$e$ & Eccentricity & &  0.36$^{+0.08}_{-0.07}$ & 0.39$^{+0.07}_{-0.07}$\\
$\omega$& Argument of periastron (deg) & &  $-$32$^{+28}_{-18}$ & $-$50$^{+16}_{-9}$\\
$i$   & Inclination (deg) & &  84.78$^{+0.43}_{-0.55}$ & 85.22$^{+0.27}_{-0.37}$\\
$\rho$   & Bulk density (g cm$^{-3}$) & &  3.13$^{+0.85}_{-0.74}$ & 3.16$^{+0.86}_{-0.73}$\\
T$_{eq}$  & Equilibrium temperature (K) & &  1119$^{+9}_{-8}$ & 1119$^{+9}_{-8}$\\
$b$  & Transit impact parameter & &  0.91$^{+0.03}_{-0.05}$ & 0.93$^{+0.02}_{-0.03}$\\
$b_S$  & Eclipse impact parameter & &  0.63$^{+0.19}_{-0.15}$ & 0.51$^{+0.12}_{-0.09}$\\
\hline
\multicolumn{5}{@{}p{15cm}@{}}{\raggedright \bedit{Note that we prefer our 2-planet model, which accounts for an additional quadratic trend in the RV data (see \S\ref{Companion}). }}\\
\end{tabular}
\end{table}





\end{document}